\documentclass[11pt]{article}

\usepackage[T1]{fontenc}
\usepackage{newtxtext}
\usepackage{newtxmath}
\usepackage[margin=1in]{geometry}
\usepackage{graphicx}
\usepackage{amsmath}
\usepackage{dsfont}
\usepackage{booktabs}
\usepackage{multirow}
\usepackage{tabularx}
\usepackage{float}
\usepackage{xcolor}
\usepackage{listings}
\usepackage[style=apa,backend=biber,natbib]{biblatex}
\usepackage{authblk}
\usepackage[colorlinks,allcolors=blue]{hyperref}

\lstdefinelanguage{Stan}{
  morekeywords={data, transformed, parameters, model, generated, quantities, int, real, vector, matrix, array, ordered, for, in, lower, upper},
  morecomment=[l]{//},
  morestring=[b]"
}
\title{Beyond Shared Thresholds: The Gaussian Process--Graded Response Model for Continuous-Time Attitudinal Dynamics}

\author[1]{Clemens Baldzuhn\thanks{Corresponding author: \href{mailto:clemens.baldzuhn@gesis.org}{clemens.baldzuhn@gesis.org}. ORCID: \href{https://orcid.org/0009-0008-1023-865X}{0009-0008-1023-865X}}}
\author[1]{Frank Mangold\thanks{\href{mailto:frank.mangold@gesis.org}{frank.mangold@gesis.org}. ORCID: \href{https://orcid.org/0000-0002-9776-3113}{0000-0002-9776-3113}}}
\author[2]{Nadja Klein\thanks{\href{mailto:nadja.klein@kit.edu}{nadja.klein@kit.edu}. ORCID: \href{https://orcid.org/0000-0002-5072-5347}{0000-0002-5072-5347}}}

\affil[1]{Department Computational Social Science, GESIS -- Leibniz Institute for the Social Sciences, Cologne, Germany}
\affil[2]{Scientific Computing Center, Karlsruhe Institute of Technology, Karlsruhe, Germany}

\date{Preprint. Submitted for publication.}

\begin{document}

\maketitle

\begin{abstract}
Longitudinal attitude research requires models that accommodate continuous latent change and the ordinal measurement of attitudes. Recent Gaussian Process Item Response Theory (GP-IRT) models address these challenges only partly because they usually impose shared response thresholds across items. We overcome this restrictive assumption and introduce the Gaussian Process--Graded Response Model (GP-GRM), combining individual-specific continuous-time Gaussian Process trajectories with item-specific thresholds for ordinal responses. Three simulation studies assess predictive performance against a shared-threshold model, recovery under structured missingness, and the effects of panel length. Item-specific thresholds improve predictive performance when category structures differ, with little cost when shared thresholds suffice. Latent trait recovery remains comparatively accurate even with shorter panels, whereas estimating individual temporal dependence requires more observations. Applications to three attitude scales from a German longitudinal panel illustrate heterogeneous measurement structures and individual trajectories. By relaxing restrictive assumptions about response-category structures across items, the GP-GRM enables continuous-time modeling of attitudinal dynamics under more flexible ordinal measurement specifications.

\end{abstract}

\section{Introduction}
Understanding how attitudes emerge, stabilize, and change over time is a central objective in psychology and the social and behavioral sciences. Attitudes toward political institutions, immigration, climate change, and other societal issues may evolve in response to societal events and changing individual circumstances. To investigate such dynamics empirically, researchers often rely on longitudinal panel surveys that repeatedly measure individuals' attitudes. Recovering latent attitude trajectories from such data raises two challenges: modeling how attitudes evolve over time, and modeling how they are measured through ordinal responses.

Regarding the first challenge, panel studies observe individuals only at discrete occasions, but the underlying latent attitudes need not evolve in discrete steps. Many longitudinal analyses are based on discrete-time models, including cross-lagged panel models \citep{hamakerCritiqueCrosslaggedPanel}. Such models are tied to predefined observation points, require additional parameters as waves increase, and offer limited flexibility for representing continuous, nonlinear dynamics between measurement occasions \citep{thomasHowCaptureReciprocal, bollenLatentCurveModels, oudContinuousTimeModeling}. Gaussian Processes \citep[GPs;][]{rasmussenGaussianProcessesMachine, nealRegressionClassificationUsing} model latent trajectories directly in continuous time. Rather than prescribing a parametric form for latent dynamics, GPs place a prior directly on trajectories, whose smoothness and temporal dependence are encoded in an adjustable covariance function. This allows nonlinear subject-specific trajectories to be modeled while naturally accommodating irregularly spaced observations and incomplete longitudinal data.

The second challenge arises from data measurement: latent attitudes and related psychological constructs are typically assessed using ordinal Likert-type response scales. Treating such responses as metric may lead to distorted inference when the assumptions underlying continuous response models are not satisfied \citep{liddellAnalyzingOrdinalData}. Item Response Theory (IRT) provides an alternative by explicitly modeling the relationship between a continuous latent trait and observed ordinal responses. Polytomous IRT models such as the Graded Response Model \citep{samejimaGeneralGradedResponse} accomplish this through threshold parameters that determine how locations on the latent trait map onto ordered response categories.

Recent methodological developments have begun to combine these two perspectives by integrating GPs with ordinal latent-variable and IRT models for longitudinal data \citep[e.g.,][]{chenLatentGaussianProcess, chenDynamicOrdinalGaussian,kangFlexibleBayesianModeling,mcbrideHierarchicalGaussianProcess}, allowing continuously evolving latent traits to be estimated while retaining an explicit probabilistic model for the ordinal responses. Building on these developments, the present paper focuses on the measurement component of GP-IRT models. Existing models have employed shared-threshold parameterizations, in which a common set of response thresholds is used across items, potentially combined with item-specific location or discrimination parameters. Such specifications resemble rating-scale formulations of polytomous IRT models \citep{andrichRatingFormulationOrdered} and reduce the number of item parameters that must be estimated.

The shared-threshold formulation nevertheless entails a substantive measurement assumption: on the latent scale, the thresholds for each item are obtained from a single common threshold vector through an item-specific shift (location) and stretch (discrimination). Consequently, the relative spacing of adjacent thresholds (the threshold configuration), and thus the relative width of the response categories, is identical across items. This assumption is reasonable when items induce similar patterns of category use, for instance when they share wording, response format, and content. In heterogeneous sets of test items, however, response distributions differ systematically with item content and formulation \citep[e.g.,][]{teneyWhatPolarizesCitizens}. Items assessing confidence in different institutions or attitudes toward distinct societal issues, for example, may differ not only in their overall location on the latent continuum but also in how respondents use the available response categories, and hence in the relative spacing of their thresholds. In longitudinal research, such measurement restrictions are particularly consequential because the reconstruction of latent trajectories depends on how observed item responses are mapped onto the underlying attitude continuum. Allowing item-specific threshold configurations thus provides a natural way to relax these restrictions while retaining the continuous-time representation of latent attitudes.

To accommodate all challenges and requirements, we propose the Gaussian Process--Graded Response Model (GP-GRM), which builds on recent GP-IRT developments by extending the ordinal measurement component through item-specific thresholds. The model retains subject-specific GP priors for continuous latent trajectories while allowing the measurement specification to accommodate item-specific category structures.

The paper makes three contributions. First, from a methodological perspective, we introduce the GP-GRM, an item-specific threshold specification within the GP-IRT framework that generalizes the measurement component while retaining the continuous-time latent process formulation. Second, from an empirical perspective, we show that the GP-GRM can be employed as a sensible default: it outperforms a shared-threshold GP-IRT specification when thresholds vary across items and loses little when they do not. Additionally, we examine the GP-GRM under conditions typical of longitudinal panel research, showing that (i) interior gaps in individual response sequences are bridged well, whereas gaps at the beginning or end of the observation period are recovered less accurately, and that (ii) latent traits can be recovered from relatively few waves, whereas individual temporal dynamics require more longitudinal information. Third, from an applied perspective, we use the model to analyze three attitude scales from the GESIS Panel.pop and show how the estimated individual trajectories reveal heterogeneous patterns of attitudinal change at the individual, population, and group level.

The remainder of the paper is structured as follows. Section~\ref{sec:methodology} introduces the GP-GRM and relates the proposed model specification to shared-threshold GP-IRT models. Section~\ref{sec:simulation_studies} presents three simulation studies evaluating the consequences of the measurement-model generalization, robustness to patterns of missingness, and recovery under varying amounts of longitudinal information. Section~\ref{sec:application} presents the empirical applications. Section~\ref{sec:discussion} discusses implications, limitations, and potential extensions of the proposed framework, and Section~\ref{sec:conclusion} concludes.

\section{Methodology}
\label{sec:methodology}

In this section, we introduce the Gaussian Process--Graded Response Model (GP-GRM). We first specify the ordinal measurement model and likelihood, followed by the GP representation of individual latent trajectories. We then relate the proposed model to a shared-threshold GP-IRT specification, describe prior distributions and model identification, and outline posterior computation. Finally, given its relevance in longitudinal panel research, we discuss estimation in the presence of missing observations.

\subsection{Measurement model and likelihood}
\label{subsec:measurement}
Let $Y_{ijt} \in \{1, \dots, C\}$ denote the ordinal response of individual $i = 1, \dots, I$ to item $j = 1, \dots, J$ at measurement occasion $t = 1, \dots, T$, collected at (possibly irregularly spaced) time points $t_1 < \dots < t_T$. It is assumed that the responses originate from an underlying continuous latent response variable $\tilde{Y}_{ijt}$, which is divided into $C$ ordered response categories by $C+1$ item-specific thresholds
\begin{equation*}
    -\infty = \tau_{0,j} < \tau_{1,j} < \dots < \tau_{C-1,j} < \tau_{C,j} = \infty.
\end{equation*}
The observed response is then determined by
\begin{equation*}
    Y_{ijt} = c \quad \Longleftrightarrow \quad \tau_{c-1,j} < \tilde{Y}_{ijt} \leq \tau_{c,j},\quad c=1,\ldots,C.
\end{equation*}
We model the latent response as $\tilde{Y}_{ijt} = \alpha_j \theta_{it} + \epsilon_{ijt}$, where $\theta_{it}$ is the latent trait of individual $i$ at occasion $t$, $\alpha_j > 0$ is an item-specific discrimination parameter, and $\epsilon_{ijt}$ follows a standard logistic distribution with cumulative distribution function $F(x) = \operatorname{logit}^{-1}(x)$.\footnote{We use the logistic distribution because it is the standard choice in IRT modeling and keeps the GP-GRM comparable with existing GP-IRT specifications \citep{mcbrideHierarchicalGaussianProcess}. Probit or (complementary) log-log links would be possible alternatives; logit and probit links typically yield very similar results \citep{burknerOrdinalRegressionModels, foxBayesianItemResponse}.} The measurement component enters the model through the linear predictor
\begin{equation*}
    \phi_{ijt,c} = \tau_{c,j} - \alpha_j \theta_{it},
\end{equation*}
such that the probabilities for observing category $c$ are
\begin{equation*}
P(Y_{ijt} = c \mid \theta_{it}, \alpha_j, \boldsymbol{\tau}_j) =
\begin{cases}
F(\phi_{ijt,1}), & c = 1, \\
F(\phi_{ijt,c}) - F(\phi_{ijt,c-1}), & 1 < c < C, \\
1 - F(\phi_{ijt,C-1}), & c = C.
\end{cases}
\end{equation*}

Let $\boldsymbol{\Theta} = (\boldsymbol{\theta}_1, \dots, \boldsymbol{\theta}_I)^\top$, where $\boldsymbol{\theta}_i = (\theta_{i1}, \dots, \theta_{iT})^\top$ denotes the latent trajectory of individual $i$. In addition, let $\boldsymbol{\alpha} = (\alpha_1, \dots, \alpha_J)^\top$, and  $\boldsymbol{\tau}=(\boldsymbol{\tau}_1^\top,\ldots,\boldsymbol{\tau}_J^\top)^\top$ collect all discrimination parameters and item-specific thresholds, respectively. Finally, let $y_{ijt}$ denote the realization of $Y_{ijt}$ and let $\boldsymbol{y} = (y_{111}, y_{112}, \dots, y_{IJT})^\top$ collect all observed responses. Conditional independence of observed responses given the latent traits and item parameters yields the likelihood
\begin{equation*}
    p_{Y}(\boldsymbol{y} \mid \boldsymbol{\Theta}, \boldsymbol{\alpha}, \boldsymbol{\tau}) = \prod_{i=1}^I \prod_{j=1}^J \prod_{t=1}^T p_{Y}(y_{ijt} \mid \theta_{it}, \alpha_j, \boldsymbol{\tau}_j).
\end{equation*}

A defining feature of the measurement model is the item-specific ordered threshold vectors $\boldsymbol{\tau}_j = (\tau_{1,j}, \dots, \tau_{C-1,j})^\top$. Together with the item-specific discrimination parameters $\alpha_j$, this yields a graded response specification \citep{burknerBayesianItemResponse, samejimaGeneralGradedResponse}. Item-specific threshold vectors allow the configuration of response-category thresholds to vary across items.

\subsection{Relationship to a Gaussian Process Rating Scale Model}
\label{sec:gp-rsm}
The GP-GRM generalizes the measurement component of the GP-IRT model proposed by \citet{mcbrideHierarchicalGaussianProcess}. Their model uses a generalized rating scale formulation \citep[RSM;][]{andrichRatingFormulationOrdered, andrichControversyRaschModel, murakiFittingPolytomousItem} in which items share a common set of category thresholds while differing in location and discrimination. We therefore refer to it as the Gaussian Process--Rating Scale Model (GP-RSM) and use it as a reference model.

The measurement model of the GP-RSM can be written as
\begin{equation*}
\phi^{\text{RSM}}_{ijt,c} = \tau_c - \alpha_j(\theta_{it} - \delta_j),
\end{equation*}
where $\tau_c$, $c=1,\ldots,C-1$, denote category thresholds shared by all items, and $\delta_j$ and $\alpha_j$, $j=1,\ldots,J$, are item-specific location and discrimination parameters, respectively. On the latent scale, threshold $c$ of item $j$ is thus located at $b_{j,c} = \delta_j+\tau_c/\alpha_j$. Thresholds can therefore differ across items by a shift and a stretch, but their relative spacing is the same for all items. The GP-RSM is the special case of the GP-GRM with $\tau_{c,j} = \tau_c + \alpha_j\delta_j$. By allowing each item its own threshold vector, the GP-GRM replaces the shared threshold configuration with item-specific ones, located at $b_{j,c} = \tau_{c,j}/\alpha_j$ on the latent scale. This flexibility matters for heterogeneous item batteries, where category use may differ across items.

Section~\ref{sec:comparison_analysis} evaluates both specifications under data-generating mechanisms corresponding to both measurement models, showing that the GP-GRM predicts better when threshold configurations differ across items and loses little when they are shared.

\subsection{Prior specifications}
\label{subsec:priors}

\subsubsection{Latent trajectories}
\label{subsec:gp}
To represent temporal dependence in individual latent traits, we place independent GP priors on the latent trajectory of each individual $i = 1, \dots, I$:
\begin{equation*}
    \boldsymbol{\theta}_i \mid \ell_i \sim \mathcal{GP}\left(0, K_{\ell_i}(t,t')\right),
\end{equation*}
where $K_{\ell_i}(t,t')$ is a kernel (covariance) function and $\ell_i$ denotes an individual-specific length-scale parameter. We collect the length scales in the vector $\boldsymbol{\ell} = (\ell_1, \dots, \ell_I)^\top$ and denote by $\boldsymbol{K}_i$ the individual-specific $T \times T$ kernel covariance matrix with entries $K_{\ell_i}(t_k, t_l)$, $k, l = 1, \dots, T$. The zero mean function centers latent trajectories around zero a priori. This fixes the location of the latent scale, which, as in conventional IRT models, is otherwise not identified \citep{foxBayesianItemResponse,sanmartinIdentificationItemResponse}.
Choosing the GP prior combined with the graded response measurement model completes the GP-GRM specification.

For the present model, we use the exponentiated quadratic kernel
\begin{equation*}
    K_{\ell_i}(t,t') = \exp\left(-\frac{(t-t')^2}{2\ell_i^2}\right),
\end{equation*}
rescaled such that the marginal variances are equal to one to avoid identifiability issues with the scale of the latent trajectories.
The exponentiated quadratic kernel yields infinitely differentiable sample functions, so it assumes that attitudes change smoothly over time, with the length scale governing how quickly they can change \citep{rasmussenGaussianProcessesMachine}. This suits the present application, in which attitudes are relatively stable dispositions that typically change gradually \citep{ansolabehereStrengthIssuesUsing}, and panel waves are spaced several months apart, so abrupt changes between waves could not be resolved anyway. However, in cases where trajectories are expected to respond to exogenous shocks or to be more volatile, rougher kernels such as the Matérn class may be preferable.

With our kernel choice, the resulting prior is stationary, as the kernel depends on $t$ and $t'$ only through their distance. The length scale $\ell_i$ governs the rate at which temporal dependence decays with distance in time. Smaller values of $\ell_i$ imply a more rapid decay of temporal dependence and allow latent traits to vary more strongly over short time intervals, whereas larger values imply stronger temporal dependence and smoother trajectories. The individual-specific length scale therefore characterizes the temporal smoothness of latent trajectories.

As recommended by \citet{betancourtRobustGaussianProcess}, we use an inverse-gamma prior for the individual length scales, $\ell_i \sim \mathcal{IG}(a, b)$, with density
\begin{equation*}
    p(\ell_i \mid a, b) = \frac{b^a}{\Gamma(a)} \ell_i^{-(a+1)} \exp\left(-\frac{b}{\ell_i}\right), \quad \ell_i, a, b \in \mathbb{R}^+.
\end{equation*}
Its light left tail regularizes against implausibly short length scales, while its heavier right tail still allows smooth trajectories. We set $a = 3$ and $b = 8$ and, as a computational safeguard, truncate $\ell_i$ to $[0.5, 12]$, mirroring the shortest wave spacing and the longest panel span in our data. The prior places little mass outside this interval, and posterior mass did not accumulate at either bound.

\subsubsection{Discrimination parameters}
Following \citet{mcbrideHierarchicalGaussianProcess}, we place independent normal priors on raw log-discriminations, $\tilde{\alpha}_j \sim \mathcal{N}(0, 0.5^2)$, and set $\alpha_j = \exp(\tilde{\alpha}_j - \bar{\tilde{\alpha}})$, where $\bar{\tilde{\alpha}} = J^{-1}\sum_{j=1}^J \tilde{\alpha}_j$. This keeps the discriminations positive and gently pulls them toward one, with more than $90\%$ of the prior mass of each $\alpha_j$ in $(0, 2)$. Restricting $\alpha_j > 0$ also fixes the orientation of the latent variables, which, like their location, is otherwise not identified. The centering additionally imposes $\prod_{j=1}^{J} \alpha_j = 1$. Although the unit marginal variance of the GP prior fixes the scale of the latent trajectories, it does so only through the prior. In initial model fits without this constraint, the discrimination parameters mixed slowly, which we attribute to the weakly identified joint scale of $\boldsymbol{\alpha}$ and $\boldsymbol{\Theta}$. The product constraint resolves this by fixing the scale exactly.

\subsubsection{Threshold parameters}
Threshold priors in graded response models are usually chosen to be noninformative, for instance truncated uniform distributions \citep[p.~38]{foxBayesianItemResponse}. Along these lines, we assign diffuse but proper Gaussian priors, $\tau_{c,j} \sim \mathcal{N}(0, 4^2)$, truncated to the ordering implied by the measurement model (Section~\ref{subsec:measurement}). Since the linear predictor $\alpha_j\theta_{it}$ has a prior scale of order one, this places little constraint on threshold locations and spacing \citep[cf.][]{johnsonOrdinalDataModeling}. Prior predictive checks confirmed that this prior adequately covers the observed response distributions (Appendix~\ref{sec:prior_pred_checks}).

\subsection{Posterior estimation and computational details}
\label{sec:estimation_priors}

Applying Bayes' rule, the joint posterior distribution is proportional to
\begin{equation*}
    p(\boldsymbol{\Theta}, \boldsymbol{\ell}, \boldsymbol{\alpha}, \boldsymbol{\tau} \mid \boldsymbol{y}) \propto p_Y(\boldsymbol{y} \mid \boldsymbol{\Theta}, \boldsymbol{\alpha}, \boldsymbol{\tau}) \, p(\boldsymbol{\Theta} \mid \boldsymbol{\ell}) \, p(\boldsymbol{\ell}) \, p(\boldsymbol{\alpha}) \, p(\boldsymbol{\tau}).
\end{equation*}
Even though the posterior is analytically intractable, we can perform Bayesian posterior estimation efficiently using Markov chain Monte Carlo (MCMC) simulations. An advantage of the Bayesian approach is the direct access to uncertainty measures for every quantity derived from the trajectories, such as individual change, population distributions, and group averages, through the MCMC samples.
Our MCMC sampler uses Hamiltonian Monte Carlo with the No-U-Turn Sampler \citep[NUTS;][]{carpenterStanProbabilisticProgramming, hoffman2014no} of Stan \citep{stan-manual} (Stan code in Appendix~\ref{suppl:stan_code}), which updates all parameters jointly. Stan enforces the ordering of the thresholds by sampling the differences between adjacent thresholds on the log scale.

The latent function values and the covariance hyperparameters are strongly coupled in the posterior \citep{murrayEllipticalSliceSampling}, which produces difficult geometries known as funnels \citep{nealSliceSampling, betancourtHamiltonianMonteCarlo}. To ease this, we use a non-centered parameterization of the latent trajectories based on the Cholesky decomposition $\boldsymbol{K}_i = \boldsymbol{L}_i \boldsymbol{L}_i^\top$ of the kernel matrix. We then sample $\boldsymbol{z}_i \sim \mathcal{N}(\boldsymbol{0}, \boldsymbol{I}_T)$ and set $\boldsymbol{\theta}_i = \boldsymbol{L}_i \boldsymbol{z}_i$. This decouples the sampled variables from the length scale that governs the covariance. A small jitter term on the diagonal of $\boldsymbol{K}_i$ keeps the decomposition numerically stable \citep{flaxmanFastHierarchicalGaussian}.

Models were fitted on a high-performance computing cluster, with one CPU core per chain.

\subsection{Missing observations}
Missing observations are a major concern in panels, where long fielding periods make item nonresponse, skipped waves, or dropout more likely \citep{struminskaya2022risk}. We assume that missingness is ignorable, that is, missing at random with missingness parameters distinct from the model parameters. The assumption is mild in this setting, as it allows missingness to depend on a respondent's observed responses at other occasions, which the GP prior links to the unobserved latent states. Under this assumption, posterior inference can be based on the observed responses alone \citep{rubin1976inference, gelman1995bayesian}.

Consequently, for the GP-GRM, latent states at occasions without observed responses enter the posterior only through the GP prior. Conditional on $\ell_i$ and the latent states at observed occasions, they follow the GP conditional distribution, whose mean is a kernel-weighted combination of these states. The GP prior thus acts as the imputation model for $\boldsymbol{\theta}_i$, and $p(\theta_{it}\mid \boldsymbol{y}_{\text{obs}})$ follows directly from the joint posterior (Appendix~\ref{sec:posterior_process}).

As the temporal distance to the nearest observed occasion grows relative to $\ell_i$, the information about $\theta_{it}$ decreases and its posterior increasingly reflects the prior. Interior gaps, bounded by observations on both sides, are therefore better informed than long gaps or gaps at the beginning or end of an individual's measurement period. Section~\ref{sec:missingness_analysis} evaluates recovery under different missingness patterns, and Section~\ref{sec:truncated_analysis} varies the number of observed panel waves.

\section{Simulation studies}
\label{sec:simulation_studies}
We conduct three simulation studies to evaluate complementary properties of the GP-GRM. Study 1 compares its predictive performance with the GP-RSM under data-generating processes corresponding to both measurement specifications. Study 2 examines recovery of latent trajectories under structured patterns of missing observations. Study 3 investigates how recovery of latent trajectories and individual-specific length scales changes with the number of observed panel waves. Across all three studies, we calibrate the data-generating processes to three empirical measurement scales from the GESIS Panel.pop \citep{ZA5665} that are subsequently analyzed in Section~\ref{sec:application}: attitudes toward the energy transition, confidence in political institutions, and attitudes toward refugees. Each scale defines one empirically calibrated simulation setting, referred to as the Energy transition, Institutions, and Refugees settings in the following. The three settings differ in the number of items (4--10), response categories (5--7), measurement occasions (12--13), and estimated item and trajectory parameters.

Specifically, we fit the GP-GRM to the complete-case subsample of each empirical scale, that is, to participants with responses on all items at all measurement occasions, and take the posterior medians as ground-truth parameters, $\boldsymbol{\Theta}^* = \operatorname{med}(\boldsymbol{\Theta} \mid \boldsymbol{y})$, $\boldsymbol{\alpha}^* = \operatorname{med}(\boldsymbol{\alpha} \mid \boldsymbol{y})$, and $\boldsymbol{\tau}^* = \operatorname{med}(\boldsymbol{\tau} \mid \boldsymbol{y})$. For the data-generating process corresponding to the GP-RSM in Study 1, we analogously fit the GP-RSM and use its posterior medians as ground truth to simulate from the model. The length scales do not enter the simulation, since responses are simulated given $\boldsymbol{\Theta}^*$. The posterior medians $\boldsymbol{\ell}^* = \operatorname{med}(\boldsymbol{\ell} \mid \boldsymbol{y})$ serve as reference values for the length-scale recovery. The settings thus allow model performance to be evaluated across different measurement designs and empirical parameter configurations.

Further details on the scales and their measurement characteristics are provided in Section~\ref{sec:data_measures} and Appendix~\ref{suppl:scales_validation}, which also reports the size of the complete-case subsamples (Table~\ref{tab:merged_descriptives}). Formal details on the employed evaluation metrics are given in Appendix~\ref{suppl:evaluation_metrics}.

\subsection{Study 1: Predictive performance of GP-GRM and GP-RSM}
\label{sec:comparison_analysis}

\begin{figure}[!htbp]
    \centering
    \includegraphics[width=0.7\linewidth]{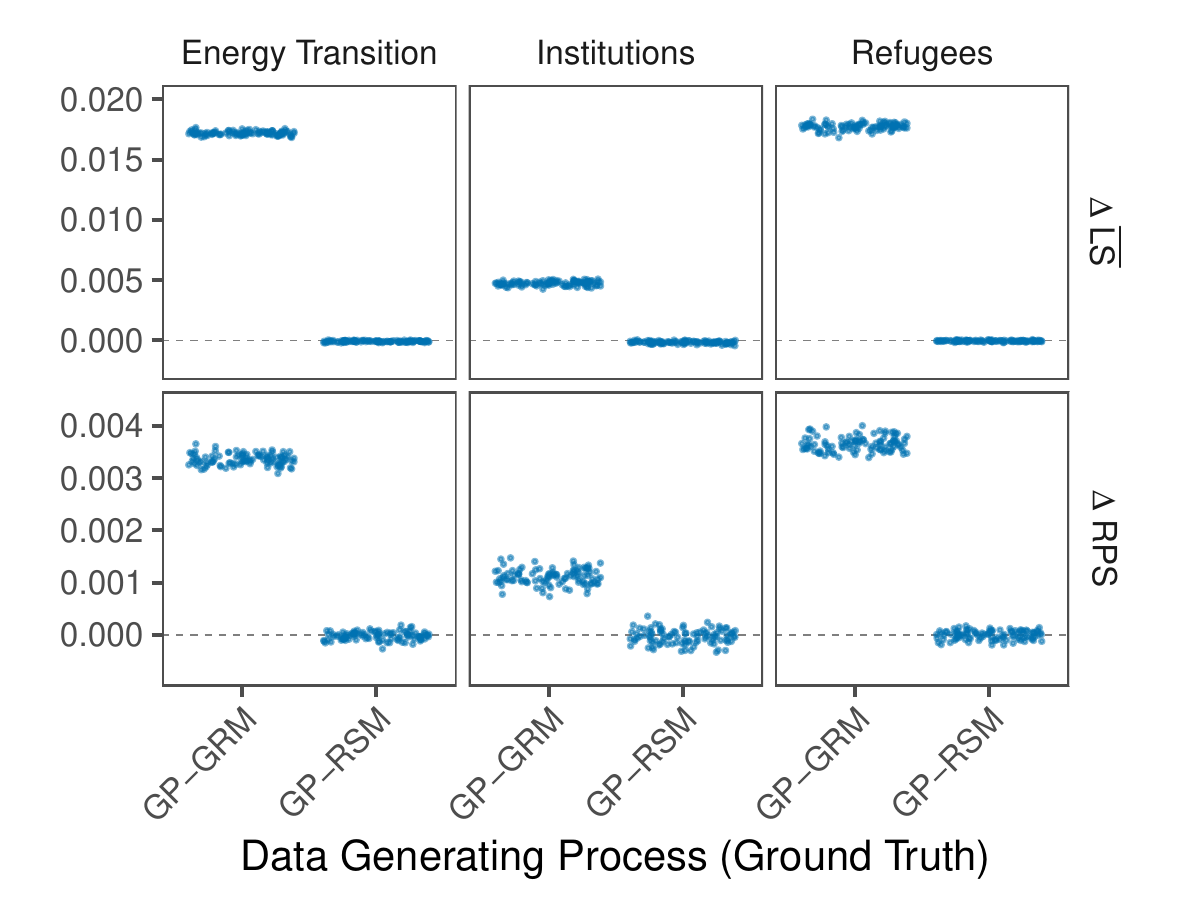}
    \caption{Differences in predictive performance between the GP-GRM and GP-RSM across the three empirically calibrated settings and two data-generating processes. Points show replicate-level differences in logarithmic score $(\Delta \text{LS})$ and ranked probability score $(\Delta \text{RPS})$, computed such that positive values indicate better predictive performance of the GP-GRM and negative values indicate better predictive performance of the GP-RSM}
    \label{fig:comparison_analysis}
\end{figure}

In Study 1, we assume $\tau_c \sim \mathcal{N}(0, 4^2)$, and for the item locations, we follow \citet{mcbrideHierarchicalGaussianProcess} and assign $\delta_j \sim \mathcal{N}(0, 1)$, constrained to sum to zero. For each of the three settings, we consider two data-generating processes (DGPs): one corresponding to the GP-GRM and one corresponding to the GP-RSM. This design quantifies the trade-off between the additional flexibility of the GP-GRM and the greater parsimony of the GP-RSM.

For each combination of setting and DGP, we generate 101 replicate datasets, fit both models to 100 of them, and evaluate predictions for the remaining holdout replicate, resulting in $2 \ (\text{models}) \times 2 \ (\text{DGPs}) \times 3 \ (\text{settings}) \times 100 \ (\text{replicates}) = 1200$ model fits. All replicates share the same ground-truth parameters, so the holdout consists of new responses for the same individuals, items, and time points. We assess predictive performance through out-of-sample density forecasting \citep[cf.][]{smithBayesianInferenceRegression}, using two proper scoring rules \citep{gneitingStrictlyProperScoring}, the logarithmic score (LS) and the ranked probability score (RPS), each averaged over held-out responses. For each fit, we compute the score difference between the GP-GRM and the GP-RSM, oriented such that positive values indicate better predictive performance of the GP-GRM.

Figure~\ref{fig:comparison_analysis} shows the resulting differences in predictive performance. When data are generated under the GP-GRM, the GP-GRM achieves better predictive performance than the GP-RSM in every replicate across all three settings and both scoring rules. The magnitude of this difference varies across settings: it is similar for the Energy transition and Refugees settings and considerably smaller for the Institutions setting. When data are generated under the GP-RSM, differences between the two models are close to zero, with no systematic advantage for either model.
The additional flexibility of the GP-GRM therefore entails negligible predictive costs when the shared-threshold specification generates the data. Overall, these results suggest that item-specific threshold configurations yield consistent predictive gains when threshold heterogeneity is present, while the costs of additional flexibility remain negligible under the shared-threshold DGPs considered here.

\subsection{Study 2: Recovery under structured missingness}
\label{sec:missingness_analysis}

\begin{figure}[!htbp]
    \centering
    \includegraphics[width=1\linewidth]{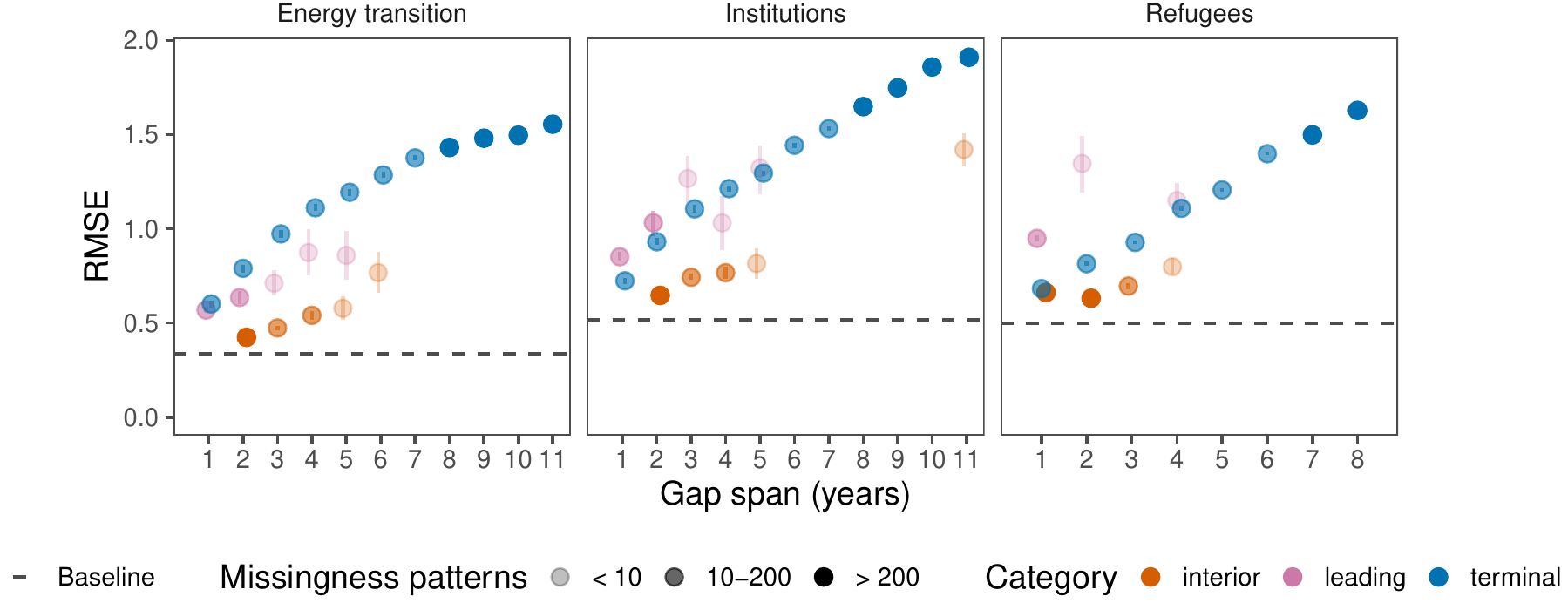}
    \caption{Recovery of latent traits across missingness categories and gap spans in the three empirically calibrated settings. Points show the RMSE between posterior medians of $\theta_{it}$ at masked occasions and their ground-truth values $\theta^{*}_{it}$, separately for interior, leading, and terminal missingness. Point opacity indicates the number of missingness patterns contributing to each condition. The dashed line shows the RMSE between posterior medians and ground-truth values in the unmasked baseline replicate. Error bars show 95\% bootstrap confidence intervals}

    \label{fig:missingness_analysis}
\end{figure}

To approximate missingness patterns encountered in the empirical panel data, we use observed wave-level missingness patterns as templates for generating masked replicate datasets.

For each setting, we first construct a binary response-availability vector for each participant in the original panel, indicating the presence or absence of responses at each measurement occasion. Missingness is defined at the wave level, as item-level nonresponse is rare. We then apply these empirical missingness patterns to a complete replicate dataset generated from the ground-truth parameters (Section~\ref{sec:simulation_studies}). Patterns are sampled with replacement, since the number of empirical missingness patterns does not match the number of individuals in the replicate dataset. We generate 100 masked replicate datasets for each of the three settings and refit the GP-GRM to each replicate.

We classify each masked cell $(i,t)$ along two dimensions. The first is its missingness category $\kappa \in \{\text{leading}, \text{interior}, \text{terminal}\}$. A cell is leading if it lies before participant $i$'s first observed wave, interior if it lies between two observed waves, and terminal if it lies after the last observed wave. The second is its gap span $g$, the time between the last observed wave before and the first observed wave after the missing episode that contains the cell. For leading gaps, the span is measured from the first panel wave to the first observed wave; for terminal gaps, from the last observed wave to the final panel wave.
We measure $g$ in years rather than in waves because the spacing of measurement occasions differs across the settings, and round it to the nearest whole year, counting spans under half a year as one year. This yields the same yearly bins for all settings and categories, and no bin is excluded. Figure~\ref{fig:illustrate_missingness} in Appendix~\ref{suppl:evaluation_metrics} illustrates both dimensions.

Recovery is evaluated by comparing posterior estimates of $\theta_{it}$ at masked occasions with their ground-truth values $\theta^{*}_{it}$ (Section~\ref{sec:simulation_studies}). We consider root mean squared error (RMSE) and correlation between posterior medians and ground-truth values, together with mean 95\% credible-interval width as a measure of posterior uncertainty. Because the number of contributing cells differs across conditions, we quantify the uncertainty in these point estimates with 95\% bootstrap confidence intervals based on 1,000 bootstrap replicates, resampling cells with replacement within each combination of missingness category and gap span. As a baseline, the same recovery metrics are computed for latent traits estimated from the unmasked replicate, evaluated over all occasions. To assess the variability of this baseline, we additionally computed RMSE for 100 complete replicate datasets in the Energy transition setting. Baseline RMSE varied only negligibly across replicates (0.327--0.345).

Additional details on the evaluation metrics are provided in Appendix~\ref{suppl:evaluation_metrics}.

Figure~\ref{fig:missingness_analysis} shows the RMSE as a function of missingness category and gap span. Across all three settings, the RMSE generally increases with the length of the missing episode. Recovery is consistently better for interior gaps than for leading and terminal gaps, reflecting the greater information available when an unobserved latent state is surrounded by observed measurement occasions. Leading and terminal missingness require inference beyond the observed portion of an individual's trajectory and consequently show larger recovery errors.

Differences between leading and terminal missingness are smaller and less consistent. In some conditions, leading gaps show somewhat larger RMSE. However, leading missingness is comparatively rare in the empirical missingness patterns, resulting in fewer contributing observations and greater uncertainty for these conditions. Results for correlation and mean credible-interval width, reported in Figure~\ref{fig:missingness_cor_ci} in Appendix~\ref{suppl:sim_figures}, show corresponding patterns. 

\subsection{Study 3: Recovery under varying numbers of panel waves}
\label{sec:truncated_analysis}

\begin{figure}[!htbp]
    \centering
    \includegraphics[width=0.8\linewidth]{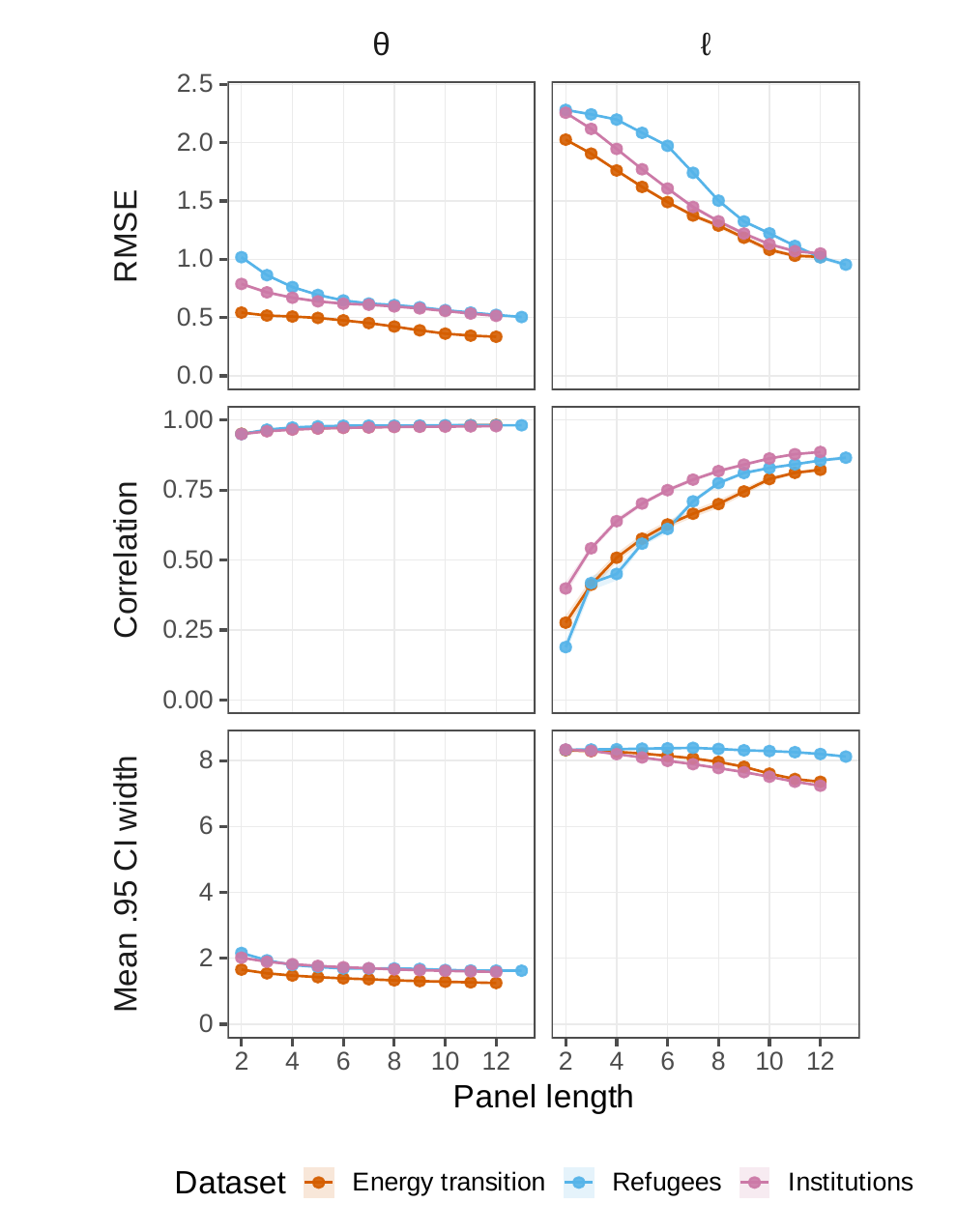}
    \caption{Recovery of latent traits and length scales across panel lengths in the three empirically calibrated settings. Rows show the RMSE and the correlation between posterior medians and reference values, and the mean width of the 95\% credible intervals. Columns show the latent traits and the length scales. All metrics are averaged over persons and over 10 replicate datasets. Shaded bands show standard deviations across replicates}
    \label{fig:truncated_analysis}
\end{figure}

For each of the three settings, we take 10 complete replicate datasets and truncate each of them to its first $2, \dots, T$ waves. We refit the GP-GRM to every truncated replicate. For each panel length, we then compute the RMSE and the correlation between posterior medians and the reference values $\theta^{*}_{it}$ and $\ell^{*}_i$, as well as the mean width of the 95\% credible intervals. All metrics are averaged over persons and over the 10 replicate datasets. Formal definitions are given in Appendix~\ref{suppl:evaluation_metrics}.

Figure~\ref{fig:truncated_analysis} shows distinct patterns for the two parameter types. Recovery of the latent traits $\theta_{it}$ depends only weakly on panel length. In the Energy transition and Institutions settings, the RMSE is already low with few waves and decreases only gradually as waves are added. In the Refugees setting, the RMSE starts higher and benefits most from the first few additional waves before leveling off at a level similar to the Institutions setting. Correlations with the ground-truth values $\theta^{*}_{it}$ remain close to one in all three settings. Mean credible-interval width also changes little.

Recovery of the length scales $\ell_i$ is more sensitive to the amount of longitudinal information. As waves are added, RMSE decreases and correlations with the reference values $\ell^{*}_i$ increase. Estimating individual temporal dependence thus benefits more from repeated observations than estimating latent traits at individual measurement occasions. Mean credible-interval width for $\ell_i$ changes little with panel length. It remains stable in the Refugees setting and decreases moderately in the Energy transition and Institutions settings. Additional waves therefore improve the posterior medians of $\ell_i$, as the RMSE and correlation show, but do not markedly narrow their posterior intervals within the panel lengths considered here.

\section{Empirical applications}
\label{sec:application}
We apply the GP-GRM to three attitude scales from the GESIS Panel.pop covering different domains of public opinion in Germany. The applications illustrate two complementary aspects of the model: its ability to accommodate item-specific measurement structures and to estimate continuous-time latent attitude trajectories. We first examine the item-specific threshold structures, that is, heterogeneity in item location, discrimination, and threshold configuration (Section~\ref{sec:item_structures}). We then analyze the estimated trajectories at the individual, population, and group levels (Section~\ref{sec:individual_dynamics}).

We fitted the GP-GRM to the analytic sample of each scale (Section~\ref{sec:data_measures}) using eight chains with 1,000 warmup and 2,000 sampling iterations each (16,000 posterior draws in total), a target acceptance rate of 0.95, and a maximum tree depth of 10. We assessed convergence and sampling efficiency using the rank-normalized $\hat{R}$ and the bulk and tail effective sample sizes (ESS) \citep{vehtariRankNormalizationFoldingLocalization}, flagging parameters with $\hat{R} \ge 1.01$ or ESS $\le 400$. No divergent transitions or maximum tree depth hits occurred. All item parameters and more than 99.9\% of the person-level parameters met the criteria; the few flagged $\theta_{it}$ and $\ell_i$ in the Refugees and Institutions scales stem from individuals with multimodal length-scale posteriors (see Appendix~\ref{suppl:estimation} and Table~\ref{tab:convergence_grm}).

\subsection{Data and measures}
\label{sec:data_measures}
We use three multi-item measures from the GESIS Panel.pop \citep{ZA5665}: attitudes toward the energy transition \citep{franzenTwoDecadesMeasuring, preisendorfer1996schone}, confidence in political institutions \citep{andersonWinningLosingPolitical, andersonPoliticalInstitutionsSatisfaction}, and attitudes toward refugees \citep{wagnerPrejudiceGrouprelatedBehavior, heitmeyer2005gruppenbezogene}. The three scales comprise ten items with five response categories measured over 12 waves (energy transition), five items with seven response categories measured over 12 waves (institutions), and four items with five response categories measured over 13 waves (refugees).

The energy-transition scale covers different aspects of energy and climate policy, including attitudes toward renewable energy, fossil fuels, nuclear power, and the economic consequences of the transition. The political-institutions scale measures confidence in the European Union, the federal government, parliament, political parties, and politicians. The refugee-attitude scale captures evaluations of migration and societal integration, including willingness to interact with refugees, compassion and sympathy toward refugees, and acceptance of refugees in one's residential environment. All items are coded such that higher values indicate higher levels of the respective construct. Descriptive scale diagnostics indicate high internal consistency (Cronbach's $\alpha = .88$--$.93$) and are consistent with a unidimensional measurement structure.

To retain respondents with incomplete longitudinal records while avoiding the boundary missingness associated with the largest recovery errors in Section~\ref{sec:missingness_analysis}, we include respondents who answered at least one item in both the first and last wave of the respective scale. Any number of interior waves may be missing. Given its low frequency and previously reported limited effects on estimation \citep{mcbrideHierarchicalGaussianProcess}, we do not separately investigate the effects of item nonresponse. Among the respondents with at least one observed wave (energy transition: $4250$; institutions: $4316$; refugees: $3268$), the resulting analytic samples comprise $1948$, $1970$, and $1909$ respondents, respectively. Item wording, descriptive statistics, scale diagnostics, and field periods are reported in Appendix~\ref{suppl:scales_validation}.

\subsection{Item-specific measurement structures}
\label{sec:item_structures}

\begin{figure}[ht]
\centering  
    \includegraphics[width=0.8\linewidth]{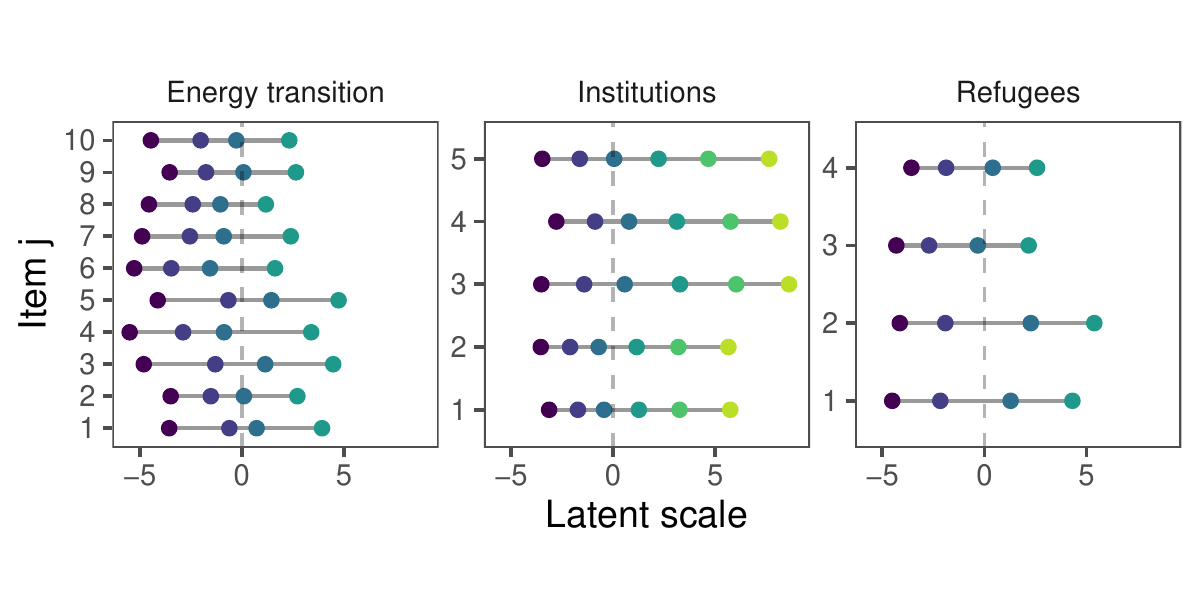}
    \caption{Item-specific threshold structures estimated by the GP-GRM for the energy-transition, institutional-confidence, and refugee-attitude scales. Points show the threshold positions of each item on the common latent scale, colored by response-category transition. Lines connect the thresholds within an item. Credible intervals are drawn but narrow}
    \label{fig:thresholds_latent}
\end{figure}

Figure~\ref{fig:thresholds_latent} displays the item-specific threshold positions $b_{j,c}$ estimated by the GP-GRM on the common latent scale (Section~\ref{sec:gp-rsm}). Unlike the raw thresholds $\tau_{c,j}$, these positions account for differences in item discrimination and therefore allow direct comparisons across items. We computed them from the posterior means of $\tau_{c,j}$ and $\alpha_j$ (per-draw computation gives nearly identical positions). The three item characteristics introduced in Section~\ref{sec:gp-rsm} are directly visible in Figure~\ref{fig:thresholds_latent}: location as the position of an item's thresholds, discrimination as their spread (lower discrimination, wider spread), and configuration as the relative spacing of adjacent thresholds (the latter distinguishing the GP-GRM from the GP-RSM).

The three scales differ markedly in their patterns of item heterogeneity. The energy-transition scale exhibits substantial variation in all three characteristics. Beyond differences in location and stretch, some items have relatively closely spaced middle categories and wider gaps toward the extremes, whereas others exhibit more evenly spaced thresholds. These differences arise in a scale covering diverse aspects of energy policy, including renewable energy, nuclear power, and economic consequences of the transition. By contrast, the confidence-in-political-institutions scale shows clear differences in item location and spread, but remarkably similar threshold configurations across institutions. The refugee-attitude scale occupies an intermediate position, with pronounced differences in location and spread but more moderate variation in threshold configuration.

These findings highlight that the relevance of item-specific threshold configurations varies across attitudinal domains. While the energy-transition scale exhibits pronounced differences in threshold configuration, the estimated configurations for confidence in political institutions are broadly consistent with the GP-RSM assumption. The GP-GRM accommodates both cases without imposing differences where the data do not support them.

\subsection{Latent attitude trajectories}
\label{sec:individual_dynamics}
\begin{figure}[!htbp]
    \centering
    \includegraphics[width=0.65\linewidth]{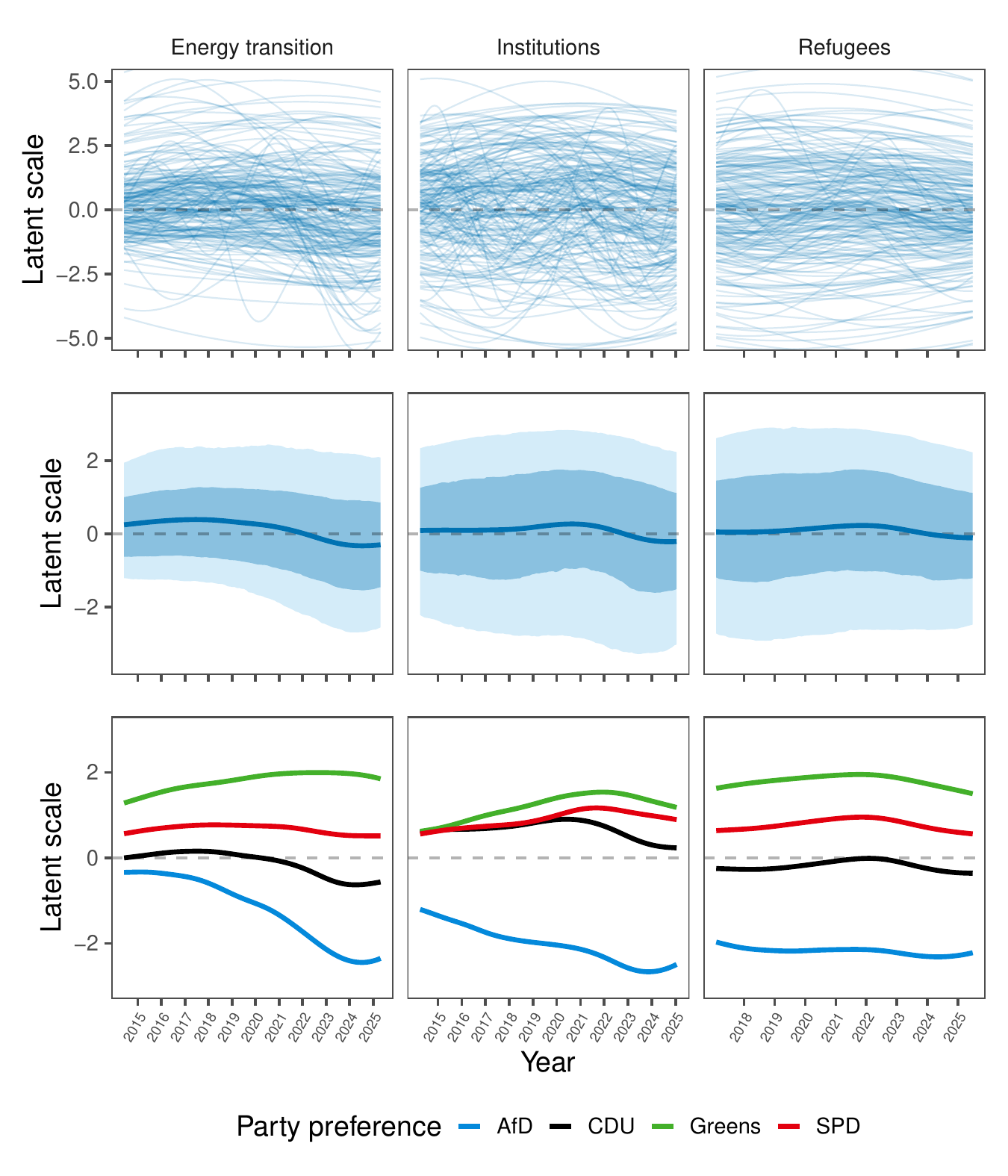}
    \caption{Latent trajectories estimated by the GP-GRM for the energy-transition, institutional-confidence, and refugee-attitude scales. All trajectories are based on the posterior mean process, interpolated between measurement occasions (Appendix~\ref{sec:posterior_process}). Upper row: trajectories of $200$ randomly selected respondents per application. Middle row: mean and $0.25$/$0.75$ and $0.10$/$0.90$ quantiles of the individual trajectories at each time point. Lower row: mean of the individual trajectories by party preference reported in the final wave}
    \label{fig:within_between_party}
\end{figure}

Figure~\ref{fig:within_between_party} illustrates how the latent attitude trajectories estimated by the GP-GRM can be examined at different levels of aggregation. The trajectories are posterior means of each respondent's latent process, evaluated both at the observed occasions and on a fine grid between them (Appendix~\ref{sec:posterior_process}). The upper row shows 200 randomly selected individual trajectories per application, revealing variation in both attitudinal levels and patterns of change. The middle row summarizes the population-level distribution of estimated trajectories over time, distinguishing shifts in central tendency from changes in between-person dispersion. The lower row presents mean trajectories grouped by respondents' party preference in the final wave, focusing on the four most frequently reported parties: Alternative für Deutschland (AfD), Christlich-Demokratische Union Deutschlands (CDU/CSU), Alliance 90/The Greens, and Sozialdemokratische Partei Deutschlands (SPD). Party preference is measured at the final wave and is not included in the estimation model. The resulting group trajectories therefore provide a retrospective, descriptive comparison rather than evidence of causal relationships.

For attitudes toward the energy transition, the population mean remains initially stable before declining during the later occasions. This aggregated pattern is accompanied by increasing between-person dispersion, particularly toward the lower end of the attitude distribution. The party-specific trajectories reveal a corresponding divergence: Greens supporters maintain comparatively high levels throughout, whereas eventual AfD supporters, initially close to CDU/CSU and SPD supporters, show a pronounced decline. Thus, the substantial differences between these groups at the final wave were not equally pronounced at the beginning of the observation period but developed over time.

For confidence in political institutions, the population mean remains largely stable. While CDU/CSU, SPD, and Greens supporters exhibit similar patterns over time, eventual AfD supporters exhibit lower estimated levels of institutional confidence from the outset and decline further during the observation period. The scale captures confidence in political institutions, not support for democratic principles, and these differences should therefore not be interpreted as direct evidence of differences in democratic commitment. Unlike the energy-transition case, the group differences are thus already apparent at the beginning and become more pronounced over time.

For attitudes toward refugees, the population mean remains largely stable, with only a modest decline toward the end of the observation period. Unlike the other two applications, the party-specific trajectories differ primarily in their levels rather than in their patterns of change. Greens supporters exhibit the highest estimated attitude levels, followed by SPD and CDU/CSU supporters, while eventual AfD supporters remain substantially lower throughout. The group differences are thus already present at the first observation and remain comparatively stable throughout the panel.

Taken together, the three applications illustrate how relatively stable population means can coexist with heterogeneous attitudinal trajectories. At the individual level, respondents differ in both their estimated attitude levels and patterns of change. At the group level, these differences take distinct forms: growing disparities in attitudes toward the energy transition, an increasing gap in institutional confidence between eventual AfD supporters and other groups, and persistent differences in attitudes toward refugees. Considering individual trajectories alongside population-level trends and group-specific developments thus reveals distinct patterns of attitudinal change and group differentiation that aggregate trends alone would obscure.

\section{Discussion}
\label{sec:discussion}

Modeling attitudinal dynamics from longitudinal survey data requires assumptions not only about how latent attitudes evolve over time, but also about how they are reflected in ordinal response values. Our results demonstrate the practical value of relaxing shared-threshold restrictions in GP-IRT models, particularly when items differ in their response-category structures.
In our simulations, the GP-GRM outperformed the GP-RSM in predicting responses when items differed in their threshold configurations, while predictive differences were generally small when items shared a common configuration. Our empirical applications reveal heterogeneous threshold structures even in established longitudinal panel scales using common response formats, although its extent varies across constructs. 

Differences in threshold configurations are pronounced for attitudes toward the energy transition, more moderate for attitudes toward refugees, and limited for confidence in political institutions. Such heterogeneity may arise from item content and wording, but also from scale construction, particularly when items capture different facets of a construct. Given that the degree of heterogeneity varies across scales and is often difficult to assess a priori, we consider the flexibility offered by item-specific thresholds a sensible default for longitudinal attitude measurement: it improves predictive performance when heterogeneity is present, with negligible loss when it is absent.

The GP prior, in turn, offers practical advantages when longitudinal information is limited: Under ignorable missingness, latent states at unobserved occasions follow directly from the posterior, so gaps can be bridged without a separate imputation step. Our simulations show that recovery depends not only on the length of a gap but also on its position. Interior gaps are recovered comparatively well because observations on both sides constrain the trajectory, whereas leading and terminal gaps produce larger errors, with RMSEs roughly one and a half to two times those of interior gaps of the same span. These findings motivate our decision to retain respondents with interior gaps while requiring observations at both the beginning and end of the panel. Beyond the first few waves, panel length has comparatively little effect on the recovery of latent trait values: additional waves reduce the RMSE only gradually, and correlations with the reference values are high even for the shortest panels. This suggests that reasonable estimates of individual latent attitudes can also be obtained from relatively short panels.

Estimating individual temporal dependence between latent attitudes is more demanding. Although individual length scales may be of substantive interest, for example in research on attitudinal inertia \citep{dennisonPsychologyPoliticalAttitudinal}, their estimation requires more information than the recovery of latent states. In Study~3, posterior medians approach the reference estimates as more waves become available, while credible intervals narrow only slightly. Moreover, in the empirical applications, some individual-specific length-scale parameters exhibit multimodal distributions. For the associated individuals, the observed responses are compatible with either a highly dynamic trajectory with a short length scale or a comparatively stable trajectory with noisy responses and a long length scale. This ambiguity illustrates the difficulty of separating temporal variation from response noise when observations are sparse. Between-person differences in estimated length scales should therefore be interpreted cautiously and should not be treated as direct measures of attitudinal inertia. As directions for future research, longer panels, hierarchical priors on length scales, or alternative covariance functions may help address these challenges.

Our applications also illustrate what individual trajectories add to substantive analyses. Across all three domains, population means change only modestly, yet party-preference groups exhibit distinct longitudinal patterns. Differences emerge over time for attitudes toward the energy transition, persist mainly in level for attitudes toward refugees, and reflect the divergence of AfD supporters for confidence in political institutions. Aggregated trends can thus conceal substantively relevant differences in attitudinal development. As the GP-GRM estimates a trajectory for each respondent, such differences can be examined for any grouping after estimation, without including group membership in the model. The analytic samples comprise respondents observed in both the first and the last wave of each scale. The ignorability assumption covers gaps between these waves but not selection into the analytic sample, so the trajectories characterize respondents who remained in the panel rather than necessarily representing the full panel sample.

The interpretation of our simulation findings also depends on how the simulations are calibrated to the empirical data. Because the reference trajectories are posterior medians from empirical model fits, they may have been smoother than typical draws from the GP prior and easier to recover, potentially leading to optimistic recovery results for short panels. Furthermore, the length scales in Study~3 are not known data-generating parameters; the study evaluates agreement with estimates from the complete-case subsample rather than recovery of true length-scale values. Simulations with trajectories generated from known GP parameters could provide a useful complement, although at the cost of moving further away from empirically calibrated settings.

Beyond these findings, the GP representation offers opportunities for studying person-specific attitudinal processes. A process-oriented perspective on psychological phenomena has long been advocated \citep{molenaarManifestoPsychologyIdiographic, vangeertProcessApproachPsychology, hamakerAnalysisIntensiveLongitudinal}. By estimating individual trajectories while pooling information across respondents for the item parameters, the GP-GRM supports person-specific analyses within a common measurement framework. This may be particularly useful for studying attitudinal development in conventional panel settings, despite their relatively limited number of measurement occasions. An additional advantage is that latent trajectories are defined in continuous time rather than only at observed survey waves. One potential application is the combination of survey-based attitude measurement with digital behavioral trace data \citep{fanExploringTemporalDynamics, mangoldNeueInfrastrukturenFuer, stierIntegratingSurveyData}, where observations often occur between survey waves. Such analyses could draw on the posterior distribution of the latent trajectories to account for measurement uncertainty, although their informativeness remains constrained by the available survey observations and the assumed temporal covariance structure.

Several extensions follow directly from the measurement and temporal assumptions of the GP-GRM. On the measurement side, hierarchical threshold specifications could borrow information across items, particularly when extreme categories are rarely observed. Such extensions would require parameterizations that preserve threshold ordering \citep{paulewiczBhsdtrPackageGeneralpurpose}. The GP-GRM assumes that item thresholds and discrimination parameters remain constant over time. Violations of this assumption could affect the interpretation of estimated latent trajectories.
On the temporal side, estimating latent attitudinal dynamics hierarchically could lead to improved predictive performance by pooling information across respondents. Furthermore, alternative kernels could encode different assumptions about how smoothly attitudes change over time. Finally, computational scalability of our current implementation is limited, as exact GP inference involves covariance factorizations whose computational cost grows cubically with the number of measurement occasions. Approximate GP representations, such as Hilbert space models \citep{riutort-mayolPracticalHilbertSpace} or nearest-neighbor GPs \citep{dattaHierarchicalNearestNeighborGaussian}, could extend the GP-GRM to panels with many more occasions, as recently applied to ordinal panel models \citep{mcbrideHierarchicalGaussianProcess}.

\section{Conclusion}
\label{sec:conclusion}

We proposed the GP-GRM, which extends the measurement component of GP-IRT models by allowing item-specific threshold configurations while retaining continuous-time latent trajectories. Our results show that this additional flexibility improves predictive performance in the presence of item heterogeneity, with comparatively small costs when a common configuration is adequate. The GP prior further enables the reconstruction of individual latent trajectories from incomplete and relatively short panels, although estimating individual temporal dependence requires more longitudinal information. By combining flexible ordinal measurement with person-specific trajectories, the GP-GRM provides a framework for examining heterogeneous attitudinal development that may remain obscured in aggregate trends.

\newpage

\paragraph{Acknowledgments}
The authors acknowledge the computational resources provided by the state of Baden-Württemberg through bwHPC. During the preparation of this work, the authors used Claude Code with Claude Opus 5.5 (Anthropic, \url{https://claude.com/product/claude-code}; May–October 2026) for writing and debugging analysis code and language editing, and ChatGPT with GPT-6.2 (OpenAI, \url{https://chatgpt.com}; September–November 2026) for editorial review and language refinement. All AI-assisted outputs were reviewed and revised as appropriate, and all analysis code was tested. The authors take full responsibility for the content of this article.

\paragraph{Data Availability Statement}
The Stan code for the GP-GRM and the GP-RSM is publicly available at \url{https://github.com/baldzuhnc/GP_GRM}, together with documented R helper functions, a worked example on a simulated panel, and a template for applying the model to other longitudinal data. The analysis scripts for the simulation studies and empirical applications are available from the corresponding author upon request. The empirical analyses use the GESIS Panel.pop Population Sample, Standard Edition (ZA5665, Version 61.0.0; \url{https://doi.org/10.4232/1.14610}). The data cannot be shared by the authors because they are provided by GESIS -- Leibniz Institute for the Social Sciences under a data use agreement that prohibits redistribution of the data and of respondent-level estimates derived from them. Researchers can obtain access to the data from GESIS by signing a data use agreement.

\paragraph{Financial Support}
This work was supported by the Deutsche Forschungsgemeinschaft (DFG, German Research Foundation) under Grant 552234710 and under Grant KL3037/1-1 (Emmy Noether Programme, awarded to Nadja Klein).

\paragraph{Competing Interests}
The authors declare none.

\paragraph{Ethical Standards}
The research meets all ethical guidelines, including adherence to the legal requirements of the study country.

\paragraph{Author Contributions}
Conceptualization: C.B. Methodology: C.B., N.K. Data curation: C.B. Visualization: C.B. Writing -- original draft: C.B. Writing -- review \& editing: C.B., F.M., N.K. All authors approved the final submitted manuscript.

\newpage

\printbibliography

\clearpage
\appendix
\counterwithin{figure}{section}
\counterwithin{table}{section}
\counterwithin{lstlisting}{section}
\numberwithin{equation}{section}

\begin{refsection}
\begin{center}
{\LARGE\bfseries Supplementary Material}
\end{center}
\bigskip

\noindent\textbf{Contents}
\begin{itemize}
    \item[\ref{sec:prior_pred_checks}] Prior predictive checks
    \item[\ref{suppl:scales_validation}] Scales and measurement characteristics
    \item[\ref{suppl:estimation}] Posterior estimation and sampling diagnostics
    \item[\ref{suppl:evaluation_metrics}] Evaluation metrics
    \item[\ref{suppl:sim_figures}] Simulation: Supplementary figures
    \item[\ref{suppl:application}] Empirical applications: Supplementary results
    \item[\ref{sec:posterior_process}] Sampling from the posterior process
    \item[\ref{suppl:stan_code}] Stan code for the GP-GRM
\end{itemize}
\clearpage

\raggedbottom
\section{Prior predictive checks}
\label{sec:prior_pred_checks}
Prior predictive checks assess whether prior choices yield sensible data distributions by simulating data from parameter values drawn from the priors themselves \citep{gabryVisualizationBayesianWorkflow}.
Formally, prior predictive simulations are obtained by drawing parameters from their prior distributions and subsequently generating responses from the likelihood $\boldsymbol{y}^{(d)} \sim p_{{Y}}(\boldsymbol{y} \mid \Psi^{(d)})$, where $\Psi^{(d)} = (\boldsymbol{\Theta}^{(d)}, \boldsymbol{\alpha}^{(d)}, \boldsymbol{\tau}^{(d)})$ collects parameter values drawn from the priors defined in Section~\ref{subsec:priors}. In practice, we let Stan generate $d=1,\dots,D=100$ post-warmup draws of $\Psi^{(d)}$ which are used to simulate from the likelihood. To facilitate interpretation, data are simulated to resemble the structure of the complete-case subsamples of the scales. The sample sizes, number of items and measurement occasions are given in Table~\ref{tab:merged_descriptives}. We gather the simulated observations in the vectors $(y_{ijt}^{(d)})_{i,j,t=1}^{I,J,T}$ and summarize them using stratified statistics: for each of the individual, wave, and item dimensions in turn, we compute the mean frequency and the $0.025$/$0.975$ quantiles of each response category, pooling across the remaining two dimensions and across all $D$ draws. For example,
\begin{equation*}
\frac{1}{D}\sum_{d=1}^{D}\frac{1}{IT}\sum_{i=1}^{I}\sum_{t=1}^{T}\mathds{1}\!\left(y^{(d)}_{i1t} = c\right), \qquad c = 1,\dots,C,
\end{equation*}
summarizes, for item $j=1$, the mean frequency of each response category $c$, pooled over all individuals, waves, and draws.

\begin{figure}[H]
    \centering
    \includegraphics[width=\linewidth]{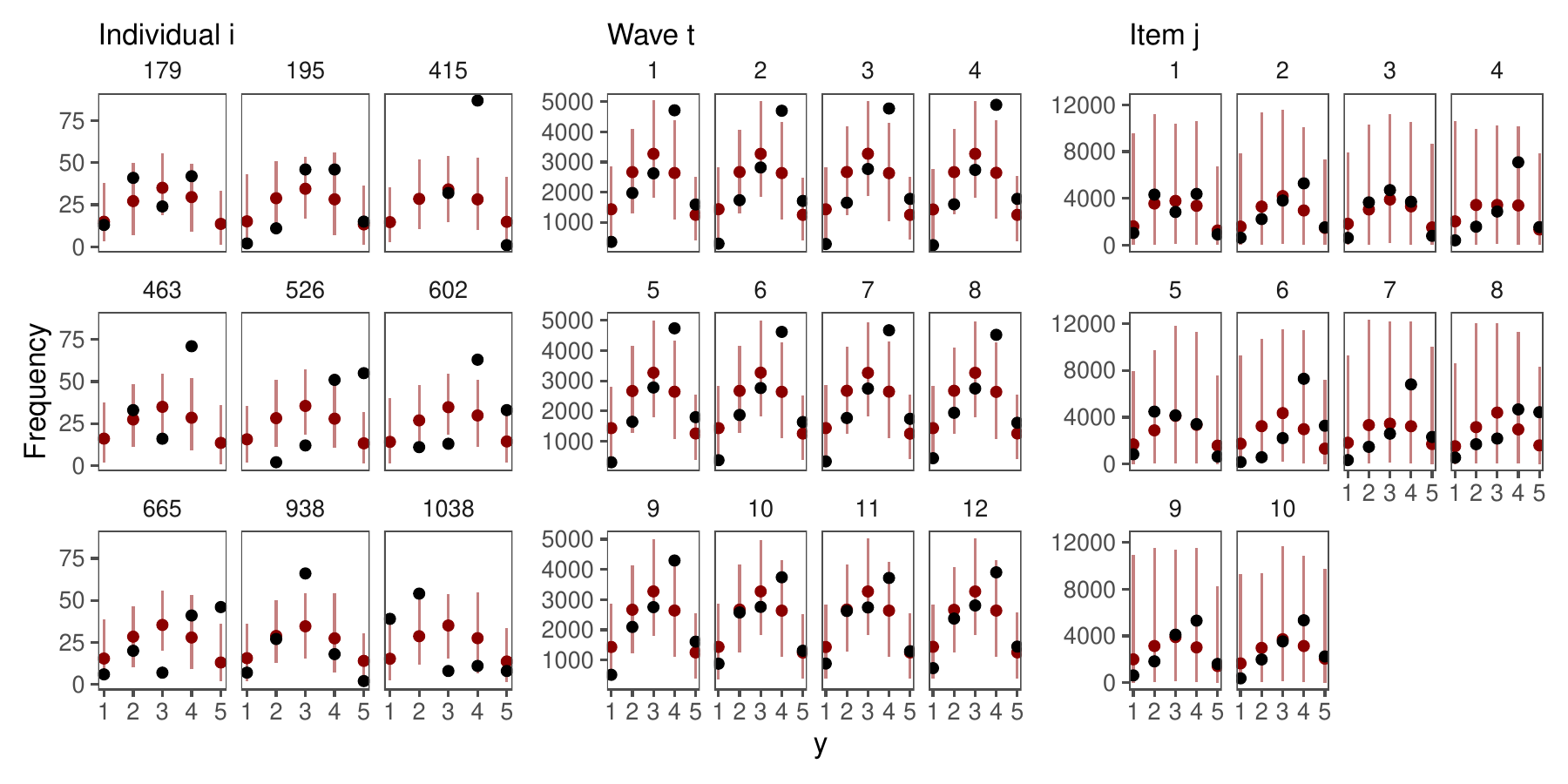}
    \caption{Prior predictive checks for the GP-GRM as specified in Section~\ref{subsec:priors}. Black: observed response frequencies in the energy-transition scale. Red: mean frequency and $0.025$/$0.975$ quantiles computed over predictive distributions from $100$ prior draws}
    \label{fig:PriorPC_ggrm}
\end{figure}

\begin{figure}[H]
    \centering
    \includegraphics[width=\linewidth]{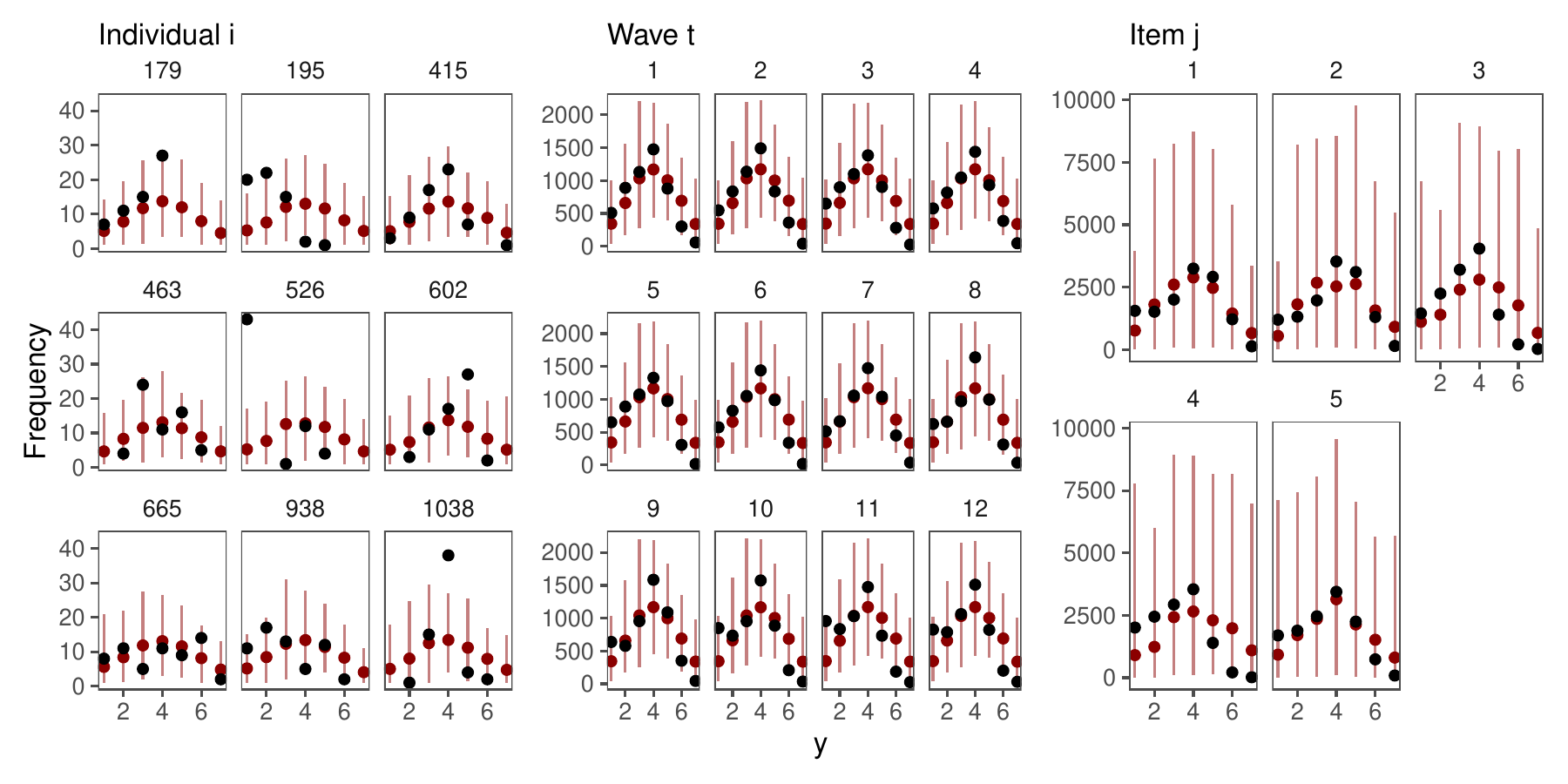}
    \caption{Prior predictive checks for the GP-GRM, confidence in political institutions. Black: observed response frequencies. Red: mean frequency and $0.025$/$0.975$ quantiles computed over predictive distributions from $100$ prior draws}
    \label{fig:PriorPC_trust}
\end{figure}

\begin{figure}[H]
    \centering
    \includegraphics[width=\linewidth]{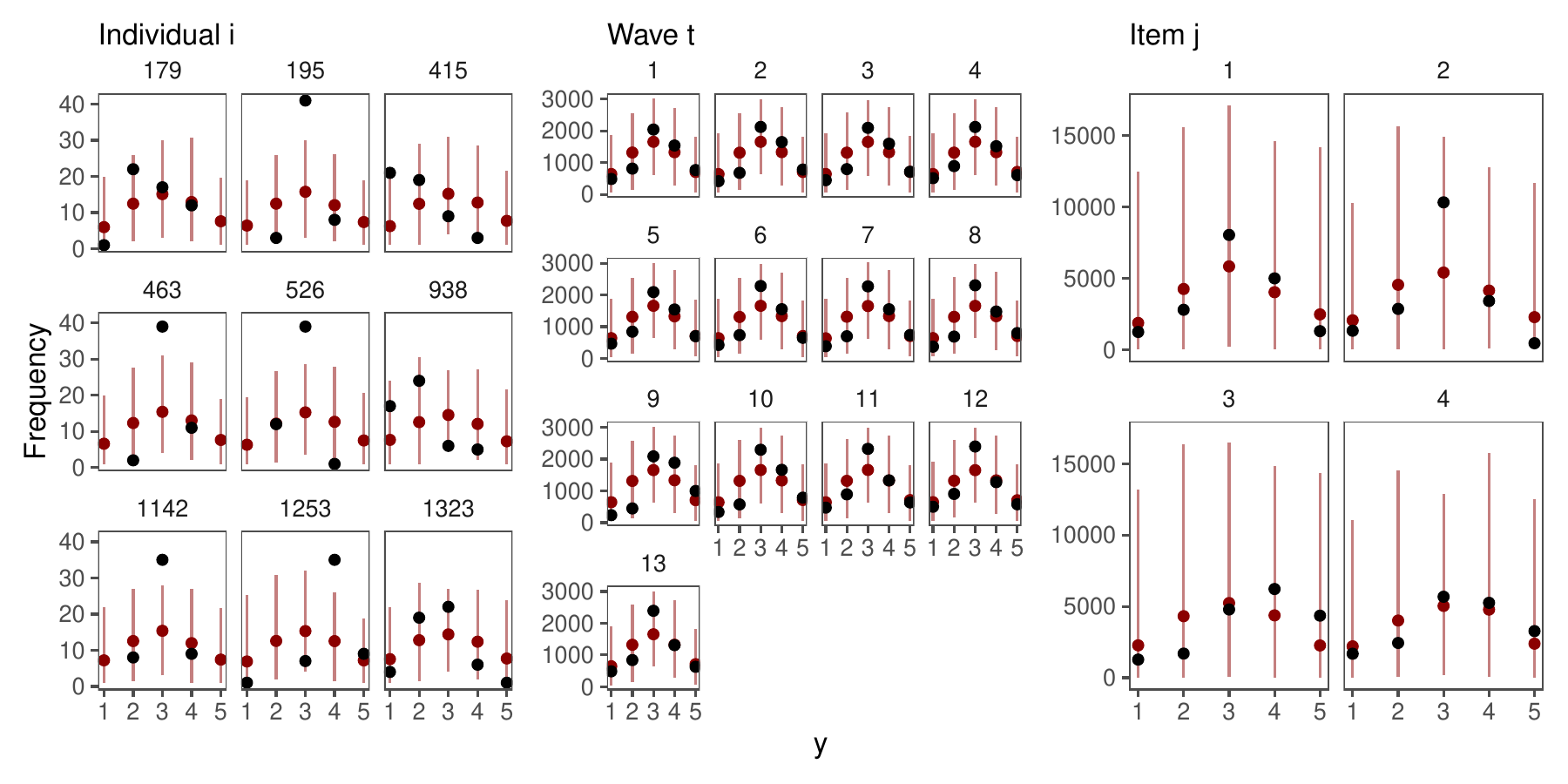}
    \caption{Prior predictive checks for the GP-GRM, attitudes toward refugees. Black: observed response frequencies. Red: mean frequency and $0.025$/$0.975$ quantiles computed over predictive distributions from $100$ prior draws}
    \label{fig:PriorPC_refugee}
\end{figure}

Figure~\ref{fig:PriorPC_ggrm} shows the prior predictive checks for the energy-transition scale. The chosen priors generate data with most mass on the middle response categories. At the item level, the prior predictive intervals are wide and cover all observed frequencies. Because the GP priors are independent across respondents and have identical marginal distributions across measurement waves, the prior predictive distribution is the same for every respondent and wave, so only the coverage of the observed range is informative at these levels. At the wave level, the intervals are narrower, since the independent item thresholds partly average out when responses are pooled over items. The observed frequencies lie within or at the boundary of these intervals. Figures~\ref{fig:PriorPC_trust} and~\ref{fig:PriorPC_refugee} show similar results for confidence in political institutions and attitudes toward refugees. In all three figures, the individual-level panels show nine randomly drawn respondents.

\clearpage

\section{Scales and measurement characteristics}
\normalsize
\label{suppl:scales_validation}
We use data from the first recruitment cohort (A1) of the GESIS Panel.pop \citep{ZA5665}. All items are coded such that higher values indicate higher levels of the respective construct.
The full sample ($I_{\text{full}}$) comprises all respondents who answered at least one wave. The complete-case subsample ($I_{\text{complete}}$) comprises respondents with responses on all items in all waves. It retains between $24\%$ and $43\%$ of the full sample and is used to obtain the reference parameters for the simulation studies (Section~\ref{sec:simulation_studies}). The interior sample ($I_{\text{interior}}$) comprises respondents observed in both the first and the last wave, with any number of interior waves missing and item nonresponse permitted. It constitutes the analytic sample of the applications (Section~\ref{sec:data_measures}). Table~\ref{tab:merged_descriptives} reports the sample sizes and descriptive statistics for the three scales. The full wording of all items is listed in Table~\ref{tab:item_wording}.

\begin{table}[!htbp]
\centering
\caption{Scale Composition, Sample Sizes, and Descriptive Statistics}
\label{tab:merged_descriptives}
\resizebox{\textwidth}{!}{%
\begin{tabular}{llcccccccc}
\toprule
\textbf{Construct / Scale}, Item & \textbf{$I_{\text{full}}$} & \textbf{$I_{\text{complete}}$} & \textbf{$I_{\text{interior}}$} & \textbf{$T$} & \textbf{$N$} & \textbf{Mean} & \textbf{St. Dev.} & \textbf{Min} & \textbf{Max} \\
\midrule
\addlinespace[0.5em]
\multicolumn{10}{l}{\textbf{Attitudes Toward Energy Transition} \textit{(Waves: bc, cc, dc, ec, fc, gc, hc, ib, jb, kb, lb, mb)}} \\
\hspace{1em}Energy saving exaggerated & \multirow{10}{*}{4250} & \multirow{10}{*}{1125} & \multirow{10}{*}{1948} & \multirow{10}{*}{12} & \multirow{10}{*}{135{,}000} & 3.69 & 0.96 & 1 & 5 \\
\hspace{1em}Future is renewables & & & & & & 3.96 & 0.83 & 1 & 5 \\
\hspace{1em}Large power plants essential & & & & & & 2.89 & 1.00 & 1 & 5 \\
\hspace{1em}Move away from fossil & & & & & & 3.03 & 0.98 & 1 & 5 \\
\hspace{1em}No alternative to transition & & & & & & 3.53 & 1.02 & 1 & 5 \\
\hspace{1em}Nuclear power dangerous & & & & & & 3.80 & 1.14 & 1 & 5 \\
\hspace{1em}Renewables insufficient & & & & & & 2.99 & 1.11 & 1 & 5 \\
\hspace{1em}Renewables make Germany independent & & & & & & 3.58 & 0.94 & 1 & 5 \\
\hspace{1em}Transition does more harm than good & & & & & & 3.36 & 1.03 & 1 & 5 \\
\hspace{1em}Transition harms industry & & & & & & 3.40 & 1.02 & 1 & 5 \\
\addlinespace[0.5em]
\multicolumn{10}{l}{\textbf{Confidence in Political Institutions} \textit{(Waves: bb, cb, db, eb, fb, gb, hb, ia, ja, ka, la, ma)}} \\
\hspace{1em}Confidence in European Union & \multirow{5}{*}{4316} & \multirow{5}{*}{1048} & \multirow{5}{*}{1970} & \multirow{5}{*}{12} & \multirow{5}{*}{62{,}880} & 3.42 & 1.47 & 1 & 7 \\
\hspace{1em}Confidence in government & & & & & & 3.68 & 1.54 & 1 & 7 \\
\hspace{1em}Confidence in parliament & & & & & & 3.84 & 1.47 & 1 & 7 \\
\hspace{1em}Confidence in political parties & & & & & & 3.19 & 1.25 & 1 & 7 \\
\hspace{1em}Confidence in politicians & & & & & & 3.05 & 1.32 & 1 & 7 \\
\addlinespace[0.5em]
\multicolumn{10}{l}{\textbf{Attitudes Toward Refugees} \textit{(Waves: ea, ec, ef, fb, ff, gc, hc, ic, jb, jc, kc, lc, mc)}} \\
\hspace{1em}Collaboration with refugees & \multirow{4}{*}{3268} & \multirow{4}{*}{1411} & \multirow{4}{*}{1909} & \multirow{4}{*}{13} & \multirow{4}{*}{73{,}372} & 3.58 & 1.15 & 1 & 5 \\
\hspace{1em}Compassion for refugees & & & & & & 3.13 & 0.98 & 1 & 5 \\
\hspace{1em}Preference residential environment & & & & & & 3.33 & 1.18 & 1 & 5 \\
\hspace{1em}Sympathy with refugees & & & & & & 2.93 & 0.85 & 1 & 5 \\
\bottomrule
\end{tabular}%
}
\begin{flushleft}
\small
\textit{Note}: $I_{\text{full}}$, $I_{\text{complete}}$, and $I_{\text{interior}}$ denote the sizes of the full sample, the complete-case subsample, and the interior sample (see text). $T$ denotes the number of waves. $N = I_{\text{complete}} \times J \times T$ denotes the number of observations in the complete-case subsample, with $J$ the number of items. Wave identifiers are GESIS Panel.pop wave codes.
\end{flushleft}
\end{table}
\normalsize

\begin{table}[htbp]

    \centering
    \footnotesize
    \caption{Full Wording of the Survey Measures}
    \begin{tabularx}{\textwidth}{l X}
        \toprule
        \textbf{Construct / Scale} & \textbf{Item} \\
        \midrule
        \textbf{Attitudes Toward Energy Transition} & All die Energiesparmaßnahmen sind letztlich übertrieben. \\
        & \textit{All those energy-saving measures are exaggerated.} \\
        \cmidrule{2-2}
        & Die Zukunft liegt in erneuerbaren Energien. \\
        & \textit{The future is in renewable energies.} \\
        \cmidrule{2-2}
        & Große Kraftwerke sind für eine sichere Stromversorgung unerlässlich. \\
        & \textit{Big power plants are essential for a reliable electricity supply.} \\
        \cmidrule{2-2}
        & Wir müssen so schnell wie möglich weg von den fossilen Brennstoffen, egal was es kostet. \\
        & \textit{We have to move away from fossil fuels as soon as possible, regardless of costs.} \\
        \cmidrule{2-2}
        & Zur Umstellung auf erneuerbare Energien gibt es keine Alternative. \\
        & \textit{There is no alternative to a switch towards renewable energies.} \\
        \cmidrule{2-2}
        & Atomkraftwerke sind eine Gefahr für uns alle. \\
        & \textit{Nuclear power plants are very dangerous for all of us.} \\
        \cmidrule{2-2}
        & Erneuerbare Energien sind einfach nicht ausreichend, um ein Industrieland zu versorgen. \\
        & \textit{Renewable energies are not sufficient to supply an industrialized country.} \\
        \cmidrule{2-2}
        & Erneuerbare Energien machen Deutschland unabhängig von anderen Ländern, weil weniger Energie importiert werden muss. \\
        & \textit{Renewable energies make Germany independent from other countries because less energy has to be imported.} \\
        \cmidrule{2-2}
        & Die sogenannte Energiewende schadet mehr als sie nutzt. \\
        & \textit{The so-called ``Energiewende'' does more harm than good.} \\
        \cmidrule{2-2}
        & Die Energiewende zerstört den Industriestandort Deutschland. \\
        & \textit{The ``Energiewende'' causes great damage to Germany as an industrial location.} \\

        \midrule
        \textbf{Confidence in Political Institutions} &  Wie sehr vertrauen Sie persönlich den folgenden öffentlichen Einrichtungen oder Personengruppen? \\
        & \textit{How much do you personally trust the following public institutions or groups?} \\
        \cmidrule{2-2}
        & Europäischen Union / \textit{European Union} \\
        & Bundesregierung / \textit{Federal government} \\
        & Bundestag / \textit{Parliament} \\
        & Politischen Parteien / \textit{Political parties} \\
        & Politikern / \textit{Politicians} \\

        \midrule

        \textbf{Attitudes Toward Refugees} & Ich kann mir vorstellen, mit Flüchtlingen zusammenzuarbeiten. \\
        & \textit{I could imagine working together with refugees.} \\
        \cmidrule{2-2}
        & Wie stark fühlen Sie mit Flüchtlingen mit? \\
        & \textit{How strongly do you sympathize with refugees?} \\
        \cmidrule{2-2}
        & Ich hätte nichts gegen Flüchtlinge in meinem Wohnumfeld. \\
        & \textit{I would not object to the presence of refugees in my residential area.} \\
        \cmidrule{2-2}
        & Wie sympathisch sind Ihnen Flüchtlinge? \\
        & \textit{How likeable are refugees to you?} \\

        \bottomrule
    \end{tabularx}
    \label{tab:item_wording}
\end{table}

\begin{figure}[!htbp]

    \centering
    \includegraphics[width=\linewidth]{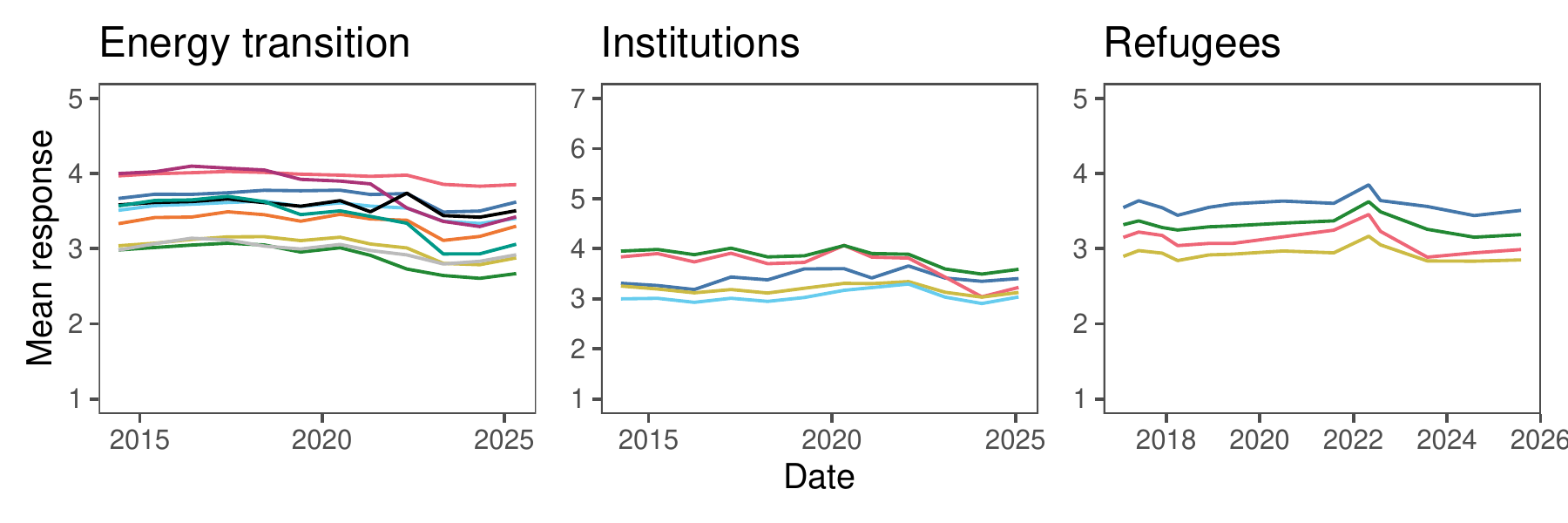}
    \caption{Mean item responses across measurement waves for the three scales, computed from the complete-case subsamples described in Table~\ref{tab:merged_descriptives}}
    \label{fig:Mean_item_response}
\end{figure}

To assess the dimensionality of the three scales, we examined internal consistency and conducted an exploratory factor analysis (EFA). Cronbach's $\alpha$ indicates high internal consistency across all three scales. The EFA results support a unidimensional structure, with first eigenvalues well above 1.0, second eigenvalues below 1.0, and factor loadings of at least 0.59 (Table~\ref{tab:scale_validation}). To complement these scale-level diagnostics, Figure~\ref{fig:Mean_item_response} provides a descriptive overview of how mean responses to individual items evolve across measurement waves, based on the complete-case subsamples. Across all three scales, item means exhibit broadly similar temporal patterns, despite differences in their average response levels. The field periods of the measurement waves are listed in Table~\ref{tab:field_periods}. These dates determine the temporal spacing of observations used in the GP models.

\begin{table}[htbp]
\caption{Internal Consistency and Exploratory Factor Analysis (EFA) Results Across Scales}
\centering
\resizebox{\textwidth}{!}{%
\small
\begin{tabular}{lccccc}

  \toprule
  \textbf{Construct / Scale} & \textbf{Cronbach's $\alpha$} & \textbf{Eig. 1} & \textbf{Eig. 2} & \textbf{Min Loading} & \textbf{Max Loading} \\
  \midrule
  Energy Transition & 0.89 & 5.12 & 0.83 & 0.59 & 0.79 \\
  Institutions      & 0.93 & 3.88 & 0.42 & 0.76 & 0.92 \\
  Refugees          & 0.88 & 3.00 & 0.48 & 0.80 & 0.83 \\
  \bottomrule
\end{tabular}%
}
\begin{flushleft} \small \textit{Note}: Eig. 1 and Eig. 2 denote the first and second eigenvalues from the exploratory factor analysis, respectively. \end{flushleft}
\label{tab:scale_validation}
\end{table}

\begin{table}[htbp]
\centering
\small
\caption{GESIS Panel.pop waves used in the three scales and their field periods}
\label{tab:field_periods}
\begin{tabular}{ll@{\hspace{2.5em}}ll}
\toprule
\textbf{Wave} & \textbf{Field period} & \textbf{Wave} & \textbf{Field period} \\
\midrule
bb & Apr--May 2014 & hc & Jul--Aug 2020 \\
bc & Jun--Jul 2014 & ia & Feb--Apr 2021 \\
cb & Apr--May 2015 & ib & May--Jul 2021 \\
cc & Jun--Jul 2015 & ic & Aug--Oct 2021 \\
db & Apr--May 2016 & ja & Feb--Apr 2022 \\
dc & Jun--Jul 2016 & jb & May--Jul 2022 \\
ea & Feb--Mar 2017 & jc & Aug--Oct 2022 \\
eb & Apr--May 2017 & ka & Feb--Apr 2023 \\
ec & Jun--Jul 2017 & kb & May--Jul 2023 \\
ef & Dec 2017--Jan 2018 & kc & Aug--Oct 2023 \\
fb & Apr--May 2018 & la & Feb--Apr 2024 \\
fc & Jun--Jul 2018 & lb & May--Jul 2024 \\
ff & Dec 2018--Jan 2019 & lc & Aug--Oct 2024 \\
gb & Apr--May 2019 & ma & Feb--Apr 2025 \\
gc & Jun--Jul 2019 & mb & May--Jul 2025 \\
hb & Apr--Jun 2020 & mc & Aug--Oct 2025 \\
\bottomrule
\end{tabular}
\begin{flushleft}
\small
\textit{Note}: Wave identifiers and field periods follow the GESIS Panel.pop documentation. The assignment of waves to scales is given in Table~\ref{tab:merged_descriptives}. Wave hb was fielded in April--May or May--June 2020.
\end{flushleft}
\end{table}

\normalsize

\clearpage

\section{Posterior estimation and sampling diagnostics}
\label{suppl:estimation}

Table~\ref{tab:convergence_grm} reports, for each parameter type, the median, worst 1\%, and worst values of $\hat{R}$ and of bulk and tail ESS, along with the number of flagged parameters ($\hat{R} \ge 1.01$ or bulk or tail ESS $\le 400$).

For Energy transition, no parameter was flagged. For Refugees, 7 of 24,817 $\theta_{it}$ and 1 of 1,909 $\ell_i$ were flagged (maximum $\hat{R} = 1.068$, minimum bulk ESS $= 90$), and for Institutions, 3 of 23,640 $\theta_{it}$ and 1 of 1,970 $\ell_i$ (maximum $\hat{R} = 1.018$, minimum bulk ESS $= 391$). The item parameters $\alpha_j$ and $\tau_{c,j}$ showed no issues in any setting ($\hat{R} \le 1.001$, bulk ESS $\ge 5{,}656$). We traced the flagged parameters back to a few individuals whose posterior distribution of $\ell_i$ is multimodal. Since these concern fewer than 0.1\% of the person-level parameters and none of the item parameters, we retained these individuals and expect negligible effects on the substantive results.

\begin{table}[htbp]
\centering
\caption{Convergence diagnostics of the GP-GRM fitted to the interior samples. Bold values exceed the thresholds $\hat{R} \ge 1.01$ or ESS $\le 400$}
\label{tab:convergence_grm}
\small
\setlength{\tabcolsep}{4.5pt}
\begin{tabular}{l r rrr rrr rrr}
\toprule
 & & \multicolumn{3}{c}{$\hat{R}$} & \multicolumn{3}{c}{Bulk ESS} & \multicolumn{3}{c}{Tail ESS} \\
\cmidrule(lr){3-5} \cmidrule(lr){6-8} \cmidrule(lr){9-11}
Parameter & Flagged & Med. & W1\% & Worst & Med. & W1\% & Worst & Med. & W1\% & Worst \\
\midrule
\multicolumn{11}{l}{\textbf{Energy transition} ($I = 1{,}948$, $T = 12$)} \\
$\theta_{it}$ & 0 / 23,376 & 1.000 & 1.001 & 1.008 & 16,140 & 6,279 & 1,049 & 13,529 & 8,906 & 1,689 \\
$\ell_i$ & 0 / 1,948 & 1.001 & 1.004 & 1.009 & 8,549 & 2,267 & 973 & 9,016 & 2,992 & 831 \\
\addlinespace
$\alpha_j$ & 0 / 10 & 1.000 & -- & 1.001 & 19,848 & -- & 18,272 & 14,118 & -- & 12,905 \\
$\tau_{c,j}$ & 0 / 40 & 1.000 & -- & 1.001 & 12,376 & -- & 8,319 & 12,484 & -- & 10,872 \\
\midrule
\multicolumn{11}{l}{\textbf{Refugees} ($I = 1{,}909$, $T = 13$)} \\
$\theta_{it}$ & 7 / 24,817 & 1.000 & 1.001 & \textbf{1.068} & 17,549 & 6,221 & \textbf{90} & 13,094 & 8,908 & \textbf{189} \\
$\ell_i$ & 1 / 1,909 & 1.001 & 1.003 & \textbf{1.065} & 8,376 & 2,356 & \textbf{113} & 7,592 & 2,961 & \textbf{298} \\
\addlinespace
$\alpha_j$ & 0 / 4 & 1.000 & -- & 1.000 & 12,137 & -- & 11,813 & 12,762 & -- & 12,309 \\
$\tau_{c,j}$ & 0 / 16 & 1.001 & -- & 1.001 & 7,602 & -- & 6,039 & 10,991 & -- & 10,280 \\
\midrule
\multicolumn{11}{l}{\textbf{Institutions} ($I = 1{,}970$, $T = 12$)} \\
$\theta_{it}$ & 3 / 23,640 & 1.000 & 1.002 & \textbf{1.013} & 15,768 & 3,916 & 429 & 12,736 & 6,234 & 980 \\
$\ell_i$ & 1 / 1,970 & 1.001 & 1.006 & \textbf{1.018} & 5,489 & 1,502 & \textbf{391} & 6,893 & 1,965 & 739 \\
\addlinespace
$\alpha_j$ & 0 / 5 & 1.000 & -- & 1.000 & 12,503 & -- & 10,952 & 12,945 & -- & 12,694 \\
$\tau_{c,j}$ & 0 / 30 & 1.001 & -- & 1.001 & 8,001 & -- & 5,656 & 11,187 & -- & 9,855 \\
\bottomrule
\end{tabular}

\vspace{2pt}
{\raggedright\footnotesize \textit{Note:} $I$ = respondents, $T$ = waves. Med.\ = median; W1\% = worst 1\% (99th percentile of $\hat{R}$, 1st percentile of ESS); Worst = maximum $\hat{R}$ or minimum ESS. Flagged = number of parameters with $\hat{R} \ge 1.01$ or bulk or tail ESS $\le 400$. W1\% is omitted for the item parameters because there are few of them.\par}
\end{table}

\section{Evaluation metrics}
\label{suppl:evaluation_metrics}

\subsection{Predictive performance of GP-GRM and GP-RSM}

\begin{description}
    \item[Posterior predictive distribution] Let $\boldsymbol{y}^{(r)}$, $r=1,\dots,R$, denote the training replicates and $\breve{\boldsymbol{y}}=(\breve{y}_{111}, \breve{y}_{112}, \dots, \breve{y}_{IJT})^\top$ the holdout replicate (Section~\ref{sec:comparison_analysis}). For the model fitted to $\boldsymbol{y}^{(r)}$, the posterior predictive distribution of $\breve{\boldsymbol{y}}$ is obtained by integrating over the whole space of unknown parameters $\Psi=(\boldsymbol{\Theta},\boldsymbol{\ell}, \boldsymbol{\alpha}, \boldsymbol{\tau})$, $p(\breve{\boldsymbol{y}}\mid \boldsymbol{y}^{(r)}) = \int p_{Y}(\breve{\boldsymbol{y}}\mid \Psi)\,p(\Psi \mid \boldsymbol{y}^{(r)}) \operatorname{d}\Psi$. In practice, this distribution is estimated via Monte Carlo integration over $D=4000$ posterior draws $\Psi^{(d)} \sim p(\Psi \mid \boldsymbol{y}^{(r)})$, $d=1,\dots,D$, such that $\hat{p}(\breve{\boldsymbol{y}}\mid \boldsymbol{y}^{(r)})= \frac{1}{D}\sum_{d=1}^{D}p_Y(\breve{\boldsymbol{y}} \mid \Psi^{(d)})$. For each draw, we additionally simulate posterior predictive responses $\breve{y}_{ijt}^{(d)} \sim p_Y(\cdot \mid \Psi^{(d)})$. For readability, we suppress the dependence of $\Psi^{(d)}$ and $\breve{y}_{ijt}^{(d)}$ on $r$.

    \item[Mean logarithmic score] The mean logarithmic score (LS) is a proper scoring rule used to evaluate how well the posterior predictive distribution predicts the held-out responses $\breve{y}_{ijt}$. For the model fitted to $\boldsymbol{y}^{(r)}$, it is defined as $\overline{\text{LS}}^{(r)} = \frac{1}{IJT}\sum_{i=1}^{I}\sum_{j=1}^{J}\sum_{t=1}^{T} \log \hat{p}(\breve{y}_{ijt}\mid \boldsymbol{y}^{(r)})$, with $\hat{p}(\breve{y}_{ijt}\mid \boldsymbol{y}^{(r)}) = \frac{1}{D}\sum_{d=1}^{D} p_Y(\breve{y}_{ijt} \mid \Psi^{(d)})$.

    \item[Mean ranked probability score] The ranked probability score (RPS) is a proper scoring rule for ordinal outcomes that compares the posterior predictive CDF with the degenerate CDF of the observed outcome. For the model fitted to $\boldsymbol{y}^{(r)}$, the score for a single held-out response is $\text{RPS}^{(r)}_{ijt} = \sum_{c=1}^{C-1}\left(\hat{F}_{\breve{y}_{ijt} \mid \boldsymbol{y}^{(r)}}(c)-\mathds{1}(\breve{y}_{ijt}\leq c)\right)^2$, where $\hat{F}_{\breve{y}_{ijt}\mid\boldsymbol{y}^{(r)}}(c) = \frac{1}{D}\sum_{d=1}^{D}\mathds{1}\left(\breve{y}_{ijt}^{(d)} \le c\right)$ denotes the empirical posterior predictive CDF. The term for $c=C$ is omitted because it is always zero. Averaging over all held-out responses gives $\overline{\text{RPS}}^{(r)} = \frac{1}{IJT}\sum_{i=1}^{I}\sum_{j=1}^{J}\sum_{t=1}^{T}\text{RPS}^{(r)}_{ijt}$. Lower RPS indicates better predictions. We therefore report $-\overline{\text{RPS}}^{(r)}$ so that, as with the LS, higher scores imply better predictive performance on held-out data \citep[for an in-depth discussion of LS and RPS, see][]{gneitingStrictlyProperScoring}.
\end{description}

\subsection{Recovery under structured missingness}

\begin{figure}[H]
    \centering
        \includegraphics[width=0.4\linewidth]{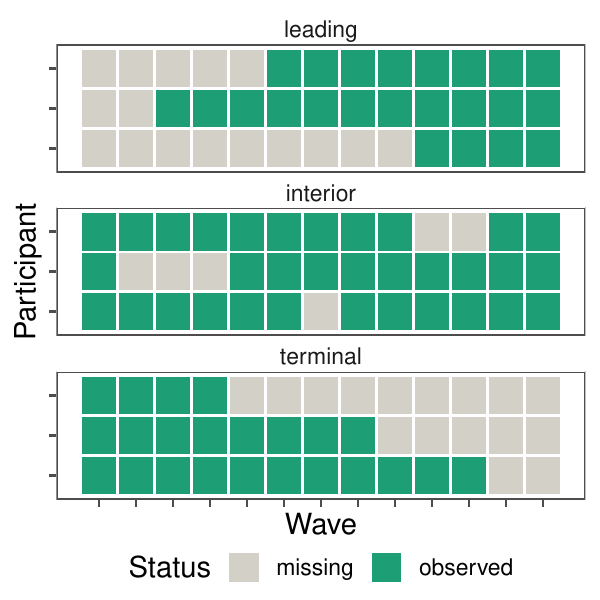}
        \caption{Illustration of the three missingness categories}
    \label{fig:illustrate_missingness}
\end{figure}

Following Section~\ref{sec:missingness_analysis}, each of the $R=100$ masked replicate datasets is obtained by applying empirical missingness patterns, sampled with replacement, to the same complete replicate dataset. Each masked cell $(i,t)$ is classified by its missingness category $\kappa \in \{\text{leading}, \text{interior}, \text{terminal}\}$ and its gap span $g$ (Figure~\ref{fig:illustrate_missingness}). The gap span is the time in years between the last observed wave before and the first observed wave after the missing episode that contains the cell. For leading and terminal episodes, the first and last waves of the panel, respectively, serve as the boundary in place of a missing observed wave. Gap spans are rounded to the nearest whole year, with halves rounded up and spans under half a year counted as one year. Let $\mathds{1}^{(r)}_{\kappa,g}(i,t)$ indicate that cell $(i,t)$ is masked in replicate $r$ and belongs to a missing episode of category $\kappa$ with gap span $g$, and let $\theta^{*}_{it}$ denote the ground-truth latent trait (Section~\ref{sec:simulation_studies}). Across the $R$ masked replicate datasets, we then compute
\begin{description}
    \item[RMSE] $\text{RMSE}_{\kappa,g} = \sqrt{\dfrac{1}{N_{\kappa,g}} \sum_{r=1}^{R}\sum_{i=1}^{I}\sum_{t=1}^{T} \mathds{1}^{(r)}_{\kappa,g}(i,t) \left(\hat{\theta}_{it}^{(r)} - \theta^{*}_{it}\right)^2}$, where $\hat{\theta}_{it}^{(r)} = \operatorname{med}(\theta_{it} \mid \boldsymbol{y}^{(r)})$ is the posterior median of $\theta_{it}$ given masked replicate $\boldsymbol{y}^{(r)}$, and $N_{\kappa,g} = \sum_{r=1}^R\sum_{i=1}^I\sum_{t=1}^T \mathds{1}^{(r)}_{\kappa,g}(i,t)$ is the number of matching cells pooled across masked replicates.
    \item[Mean 95\% credible-interval width] $\overline{W}_{\kappa,g} = \dfrac{1}{N_{\kappa,g}} \sum_{r=1}^{R}\sum_{i=1}^{I}\sum_{t=1}^{T} \mathds{1}^{(r)}_{\kappa,g}(i,t) \left(b_{it}^{U, (r)} - b_{it}^{L, (r)}\right)$, where $b_{it}^{L, (r)}$ and $b_{it}^{U, (r)}$ denote the $0.025$ and $0.975$ quantiles of the marginal posterior distribution of $\theta_{it}$ for masked replicate $r$, respectively.
    \item[Correlation] $\text{Cor}_{\kappa,g} = \text{Cor}\!\left(\hat{\boldsymbol{\theta}}_{\kappa,g}, \boldsymbol{\theta}^{*}_{\kappa,g}\right)$, where $\hat{\boldsymbol{\theta}}_{\kappa,g}$ and $\boldsymbol{\theta}^{*}_{\kappa,g}$ collect the posterior medians and ground-truth values, respectively, of all cells $(i,t,r)$ with $\mathds{1}^{(r)}_{\kappa,g}(i,t)=1$, pooled across $i=1,\dots,I$, $t=1,\dots,T$, and $r=1,\dots,R$.
\end{description}

\subsection{Recovery under varying numbers of panel waves}

Following Section~\ref{sec:truncated_analysis}, each of $R=10$ complete replicate datasets is truncated to panel lengths $T' = 2, \dots, T$ by retaining the waves $t \le T'$. Reference values are the ground-truth latent traits $\theta^{*}_{it}$ and the empirically estimated reference length scales $\ell^{*}_i$ (Section~\ref{sec:simulation_studies}). For each panel length $T'$, we compute
\begin{description}
    \item[RMSE] $\overline{\text{RMSE}}_{\ell,T'} = \frac{1}{R}\sum_{r=1}^{R} \text{RMSE}_{\ell,T'}^{(r)}$ with $\text{RMSE}_{\ell,T'}^{(r)} = \sqrt{\frac{1}{I} \sum_{i=1}^{I} (\hat{\ell}_{i,T'}^{(r)} - \ell^{*}_i)^2}$, where $\hat{\ell}_{i,T'}^{(r)} = \operatorname{med}(\ell_i \mid \boldsymbol{y}^{(r)}_{T'})$ is the posterior median of $\ell_i$ given replicate $\boldsymbol{y}^{(r)}_{T'}$, that is, replicate $r$ truncated to $T'$ waves. Analogously, $\overline{\text{RMSE}}_{\theta,T'} = \frac{1}{R}\sum_{r=1}^{R} \text{RMSE}_{\theta,T'}^{(r)}$ with $\text{RMSE}_{\theta,T'}^{(r)} = \sqrt{\dfrac{1}{IT'} \sum_{i=1}^{I}\sum_{t=1}^{T'} (\hat{\theta}_{it,T'}^{(r)} - \theta^{*}_{it})^2}$, where $\hat{\theta}_{it,T'}^{(r)} = \operatorname{med}(\theta_{it} \mid \boldsymbol{y}^{(r)}_{T'})$ is the corresponding posterior median of $\theta_{it}$.
    \item[Mean 95\% credible-interval width] $\overline{W}_{\ell,T'} = \frac{1}{R}\sum_{r=1}^{R} W_{\ell,T'}^{(r)}$ with $W_{\ell,T'}^{(r)} = \frac{1}{I} \sum_{i=1}^{I} (b_{i,T'}^{U, (r)} - b_{i,T'}^{L, (r)})$, where $b_{i,T'}^{L, (r)}$ and $b_{i,T'}^{U, (r)}$ denote the $0.025$ and $0.975$ quantiles of the marginal posterior distribution of $\ell_i$ for replicate $r$ truncated to $T'$ waves. Analogously, $\overline{W}_{\theta,T'} = \frac{1}{R}\sum_{r=1}^{R} W_{\theta,T'}^{(r)}$ with $W_{\theta,T'}^{(r)} = \dfrac{1}{IT'} \sum_{i=1}^{I}\sum_{t=1}^{T'} (b_{it,T'}^{U, (r)} - b_{it,T'}^{L, (r)})$, where $b_{it,T'}^{L, (r)}$ and $b_{it,T'}^{U, (r)}$ denote the corresponding quantiles for $\theta_{it}$.
    \item[Correlation] $\overline{\text{Cor}}_{\ell,T'} = \frac{1}{R}\sum_{r=1}^{R} \text{Cor}_{\ell,T'}^{(r)}$ with $\text{Cor}_{\ell,T'}^{(r)} = \text{Cor}(\hat{\boldsymbol{\ell}}_{T'}^{(r)}, \boldsymbol{\ell}^{*})$, where $\hat{\boldsymbol{\ell}}_{T'}^{(r)} = (\hat{\ell}_{1,T'}^{(r)}, \dots, \hat{\ell}_{I,T'}^{(r)})^\top$ and $\boldsymbol{\ell}^{*} = (\ell^{*}_1, \dots, \ell^{*}_I)^\top$. Analogously, $\overline{\text{Cor}}_{\theta,T'} = \frac{1}{R}\sum_{r=1}^{R} \text{Cor}_{\theta,T'}^{(r)}$ with $\text{Cor}_{\theta,T'}^{(r)} = \text{Cor}(\hat{\boldsymbol{\theta}}_{T'}^{(r)}, \boldsymbol{\theta}^{*}_{T'})$, where $\hat{\boldsymbol{\theta}}_{T'}^{(r)}$ and $\boldsymbol{\theta}^{*}_{T'}$ collect the posterior medians and ground-truth values, respectively, of all cells $(i,t)$ with $i=1,\dots,I$ and $t=1,\dots,T'$.
\end{description}

\section{Simulation: Supplementary figures}
\label{suppl:sim_figures}

\begin{figure}[H]
    \centering
    \includegraphics[width=0.9\linewidth]{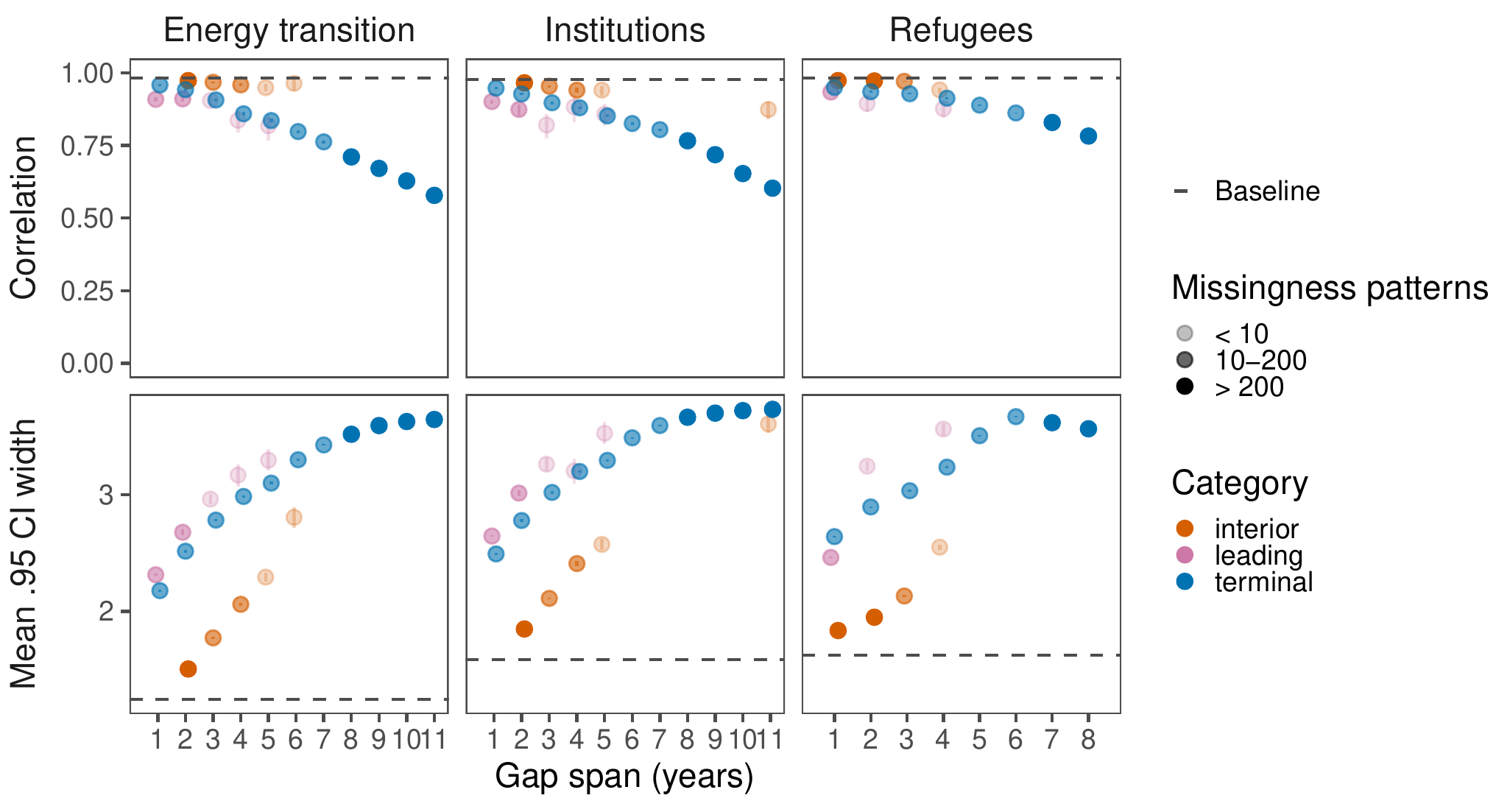}
    \caption{Correlation and mean $95\%$ credible-interval width for latent traits across missingness categories and gap spans in the three empirically calibrated settings. Upper row: correlation between posterior medians of $\theta_{it}$ at masked occasions and their ground-truth values $\theta^{*}_{it}$. Lower row: mean width of the $95\%$ credible intervals of $\theta_{it}$ at masked occasions. Point opacity indicates the number of missingness patterns contributing to each condition. Dashed lines show the corresponding values for the unmasked baseline replicate (Section~\ref{sec:missingness_analysis})}
    \label{fig:missingness_cor_ci}
\end{figure}

\section{Empirical applications: Supplementary results}
\label{suppl:application}
\subsection{Posterior estimates of the model parameters}
\label{suppl:item_posteriors}
\begin{figure}[H]
    \centering
    \includegraphics[width=\linewidth]{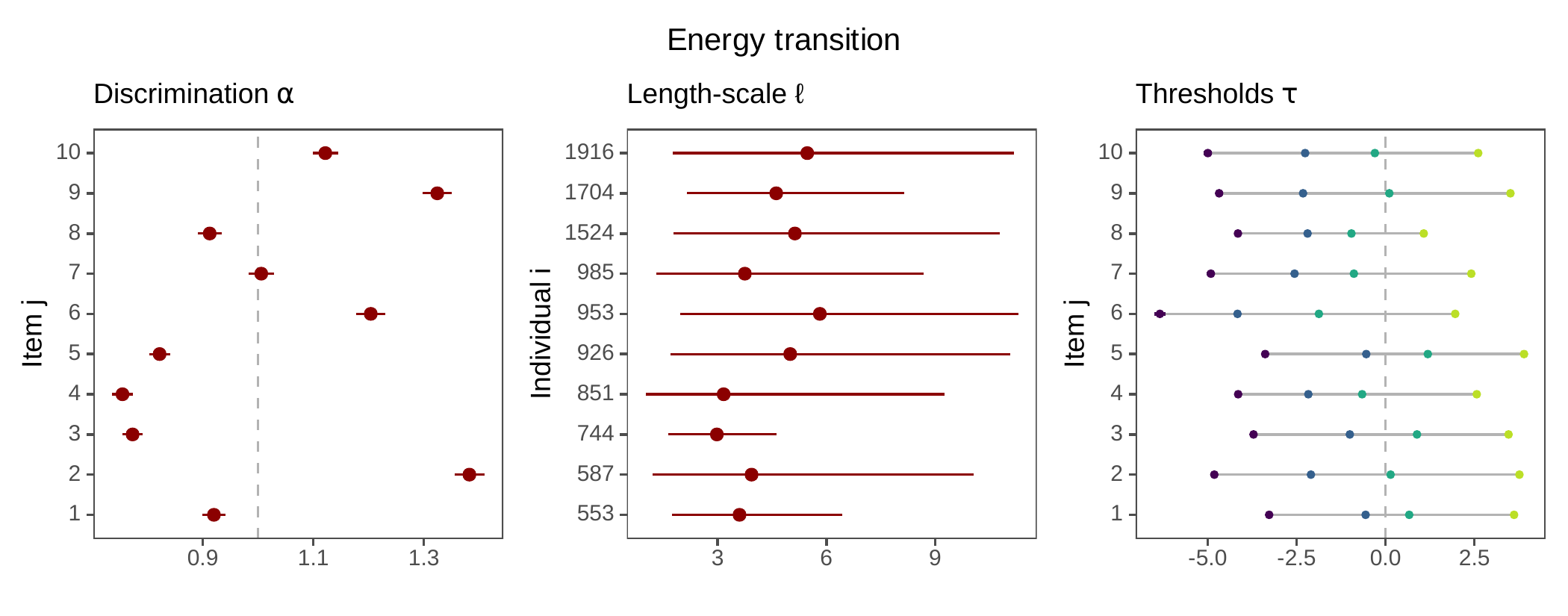}
    \caption{Raw posterior estimates of the GP-GRM parameters for the energy-transition scale: discrimination parameters $\alpha_j$, length scales $\ell_i$ for selected respondents, and thresholds $\tau_{c,j}$ on their original (unscaled) metric. Points show posterior means, and horizontal lines represent $95\%$ credible intervals}
    \label{fig:energy_parameters}
\end{figure}

\begin{figure}[H]
    \centering
    \includegraphics[width=\linewidth]{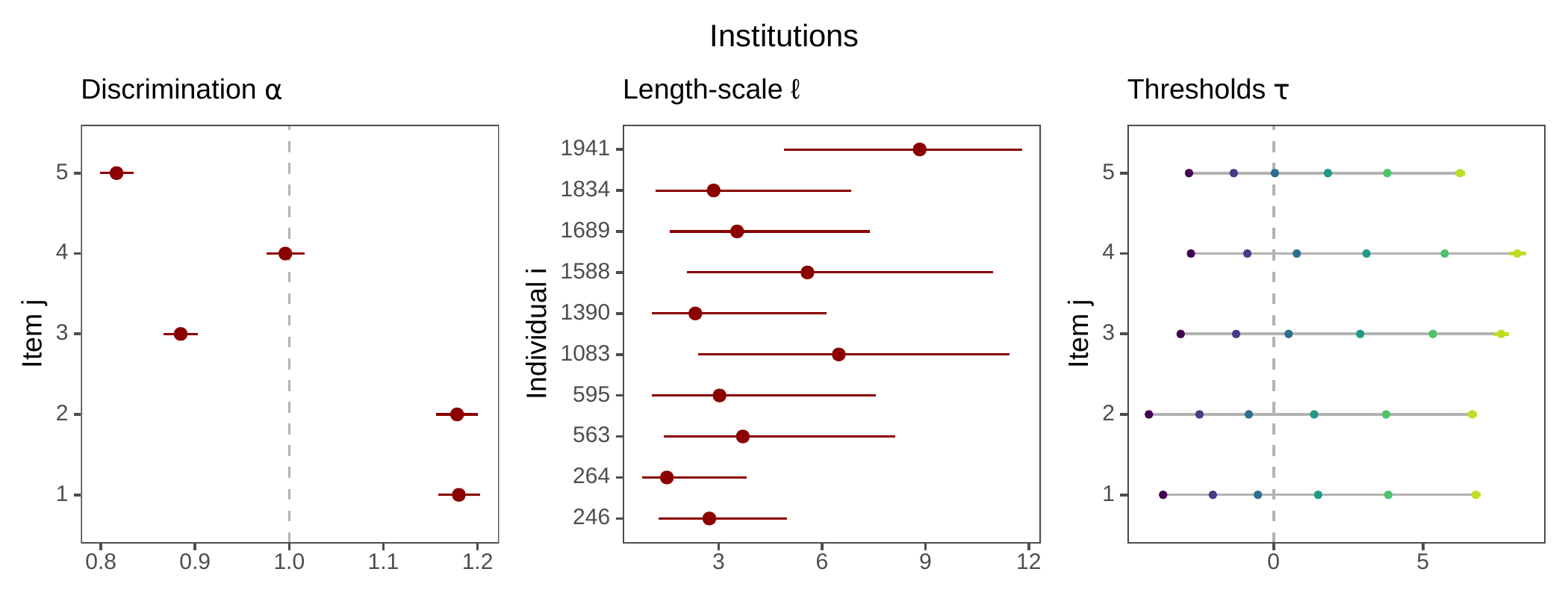}
    \caption{Raw posterior estimates of the GP-GRM parameters for the institutional-confidence scale: discrimination parameters $\alpha_j$, length scales $\ell_i$ for selected respondents, and thresholds $\tau_{c,j}$ on their original (unscaled) metric. Points show posterior means, and horizontal lines represent $95\%$ credible intervals}
    \label{fig:trust_parameters}
\end{figure}

\begin{figure}[H]
    \centering
    \includegraphics[width=\linewidth]{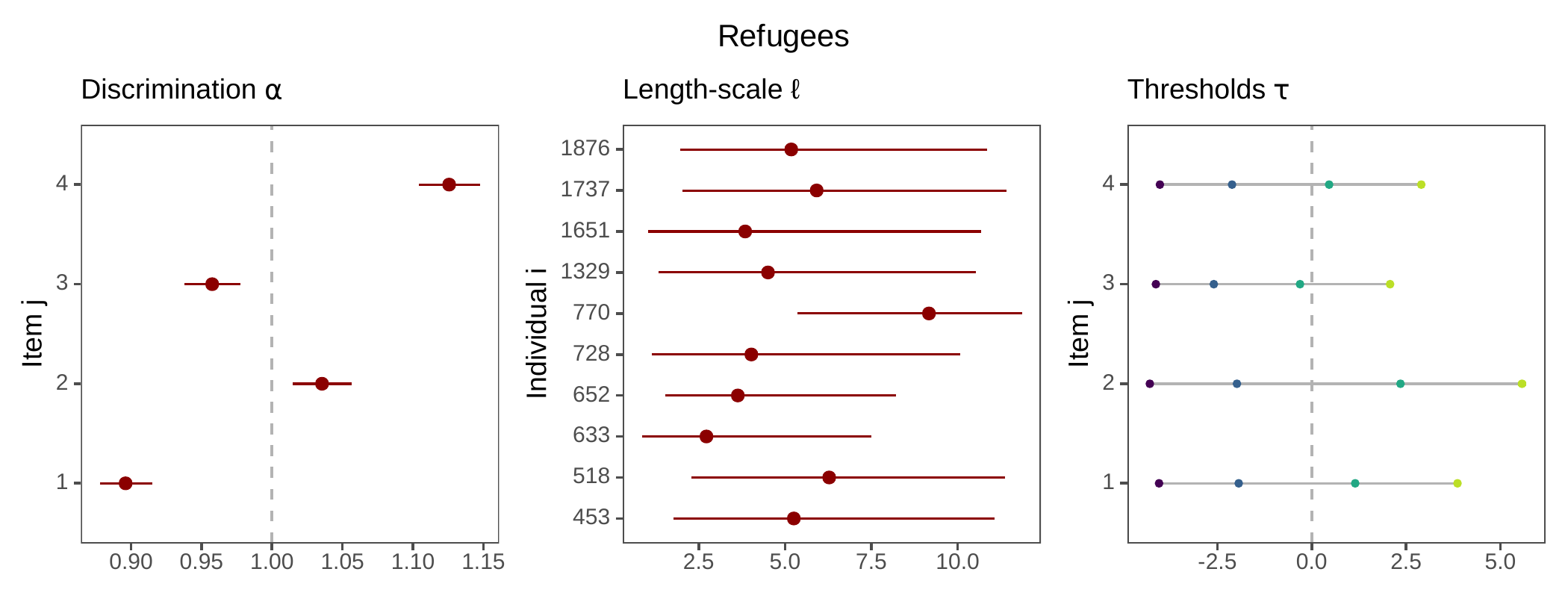}
    \caption{Raw posterior estimates of the GP-GRM parameters for the refugee-attitude scale: discrimination parameters $\alpha_j$, length scales $\ell_i$ for selected respondents, and thresholds $\tau_{c,j}$ on their original (unscaled) metric. Points show posterior means, and horizontal lines represent $95\%$ credible intervals}
    \label{fig:refugee_parameters}
\end{figure}

Figures~\ref{fig:energy_parameters}--\ref{fig:refugee_parameters} report the raw posterior estimates of the discrimination parameters $\alpha_j$, the length scales $\ell_i$, and the thresholds $\tau_{c,j}$ for the three applications, as posterior means with $95\%$ credible intervals. In contrast to Figure~\ref{fig:thresholds_latent}, the thresholds are shown on their original metric, that is, before division by $\alpha_j$.

\subsection{Posterior trajectories with uncertainty}
\label{suppl:trajectories_uncertainty}
\begin{figure}[H]
    \centering
    \includegraphics[width=0.7\linewidth]{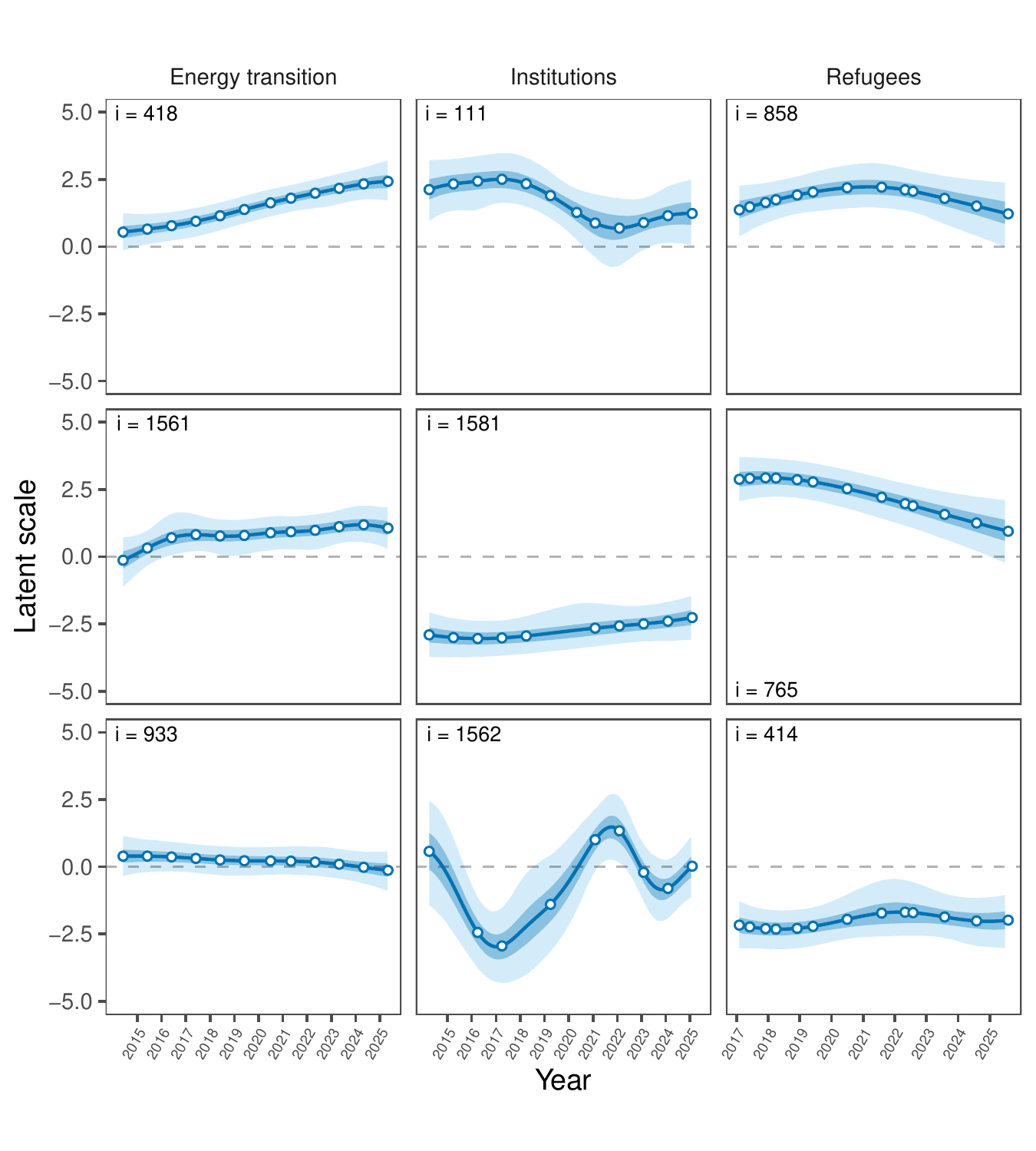}
    \caption{Posterior latent trajectories for three randomly sampled respondents per scale. Lines show posterior means. Dark bands indicate pointwise $50\%$ credible intervals (0.25--0.75 posterior quantiles), and light bands indicate pointwise $95\%$ credible intervals (0.025--0.975 posterior quantiles). Dots indicate the respondent's measurement occasions}
    \label{fig:trajectories_uncertainty}
\end{figure}

Figure~\ref{fig:trajectories_uncertainty} shows posterior latent trajectories for three randomly sampled respondents per scale from the analytic (interior) samples. The trajectories are obtained by sampling from the joint posterior of each respondent's latent process at the observed measurement occasions and on a fine grid of unobserved time points (Section~\ref{sec:posterior_process}). In contrast to Figure~\ref{fig:within_between_party}, which shows posterior means only, the figure also displays pointwise posterior uncertainty.

\clearpage

\section{Sampling from the posterior process}
\label{sec:posterior_process}

A key advantage of the GP formulation is that it yields estimates of the latent
trajectories not only at the observed time points $t_1,\dots,t_T$, but also at
arbitrary unobserved time points $t^*_1,\dots,t^*_{T^*}$. This enables nonlinear
interpolation between panel waves, which in the GP framework amounts to prediction
from noise-free function values \citep[Section~2.2]{rasmussenGaussianProcessesMachine}.
Conditional on each posterior draw of $\boldsymbol{\theta}_i$, the latent traits at the
observed time points are treated as known. The measurement error of the ordinal responses
enters the prediction only through the posterior uncertainty of $\boldsymbol{\theta}_i$. Let
$\boldsymbol{\theta}^*_i = (\theta^*_{i1},\dots,\theta^*_{iT^*})^\top$ denote the latent
traits of individual $i$ at the unobserved time points. Under the GP prior,
$\boldsymbol{\theta}_i$ and $\boldsymbol{\theta}^*_i$ are jointly Gaussian,
\begin{equation*}
\begin{pmatrix} \boldsymbol{\theta}_i \\ \boldsymbol{\theta}^*_i \end{pmatrix} \Bigm| \ell_i
\sim \mathcal{N}\left(\boldsymbol{0},
\begin{pmatrix} \boldsymbol{K}_i & \boldsymbol{K}^*_i \\ \boldsymbol{K}^{*\top}_i & \boldsymbol{K}^{**}_i \end{pmatrix}\right),
\end{equation*}
where $\boldsymbol{K}^*_i \in \mathbb{R}^{T\times T^*}$ and
$\boldsymbol{K}^{**}_i \in \mathbb{R}^{T^*\times T^*}$ denote the individual-specific
kernel matrices containing the covariances between observed and unobserved time points
and among unobserved time points, respectively, with entries
$(\boldsymbol{K}^*_i)_{km} = K_{\ell_i}(t_k, t^*_m)$ and
$(\boldsymbol{K}^{**}_i)_{mn} = K_{\ell_i}(t^*_m, t^*_n)$, for $k = 1,\dots,T$ and
$m,n = 1,\dots,T^*$. Since the item responses depend only on $\boldsymbol{\theta}_i$,
$\boldsymbol{\theta}^*_i$ is conditionally independent of the data given
$\boldsymbol{\theta}_i$ and $\ell_i$, and its conditional distribution follows from
standard properties of the multivariate normal distribution as
\begin{equation*}
\boldsymbol{\theta}^*_i \mid \boldsymbol{\theta}_i, \ell_i \sim \mathcal{N}\left(
\boldsymbol{K}^{*\top}_i\boldsymbol{K}_i^{-1}\boldsymbol{\theta}_i,\;
\boldsymbol{K}^{**}_i - \boldsymbol{K}^{*\top}_i\boldsymbol{K}_i^{-1}\boldsymbol{K}^*_i\right).
\end{equation*}
Drawing $\boldsymbol{\theta}^*_i$ from this distribution for each posterior draw of
$(\boldsymbol{\theta}_i, \ell_i)$ yields samples from the posterior
$p(\boldsymbol{\theta}^*_i \mid \boldsymbol{y})$ \citep[eq.~2.19]{rasmussenGaussianProcessesMachine}.
This is considerably more efficient than including the unobserved time points directly in the model, which would enlarge the parameter space and require factorizing $(T+T^*)\times(T+T^*)$ kernel matrices at each gradient evaluation of the sampler \citep{stan_users_guide_gp}.

The same argument applies to panel waves at which individual $i$ provided no
responses. Let $\mathcal{O}_i$ and $\mathcal{M}_i$ denote the waves with and
without observed responses, and partition $\boldsymbol{\theta}_i$ and
$\boldsymbol{K}_i$ accordingly. Since $\boldsymbol{\theta}_{i\mathcal{M}}$ does
not enter the likelihood, replacing $\boldsymbol{\theta}_i$ by
$\boldsymbol{\theta}_{i\mathcal{O}}$ and $\boldsymbol{\theta}^*_i$ by
$\boldsymbol{\theta}_{i\mathcal{M}}$ above gives
\begin{equation*}
\boldsymbol{\theta}_{i\mathcal{M}} \mid \boldsymbol{\theta}_{i\mathcal{O}}, \ell_i \sim \mathcal{N}\left(
\boldsymbol{K}_{i,\mathcal{M}\mathcal{O}}\boldsymbol{K}_{i,\mathcal{O}\mathcal{O}}^{-1}\boldsymbol{\theta}_{i\mathcal{O}},\;
\boldsymbol{K}_{i,\mathcal{M}\mathcal{M}} - \boldsymbol{K}_{i,\mathcal{M}\mathcal{O}}\boldsymbol{K}_{i,\mathcal{O}\mathcal{O}}^{-1}\boldsymbol{K}_{i,\mathcal{O}\mathcal{M}}\right).
\end{equation*}
Because $\boldsymbol{\theta}_{i\mathcal{M}}$ is part of the sampled trajectory
$\boldsymbol{\theta}_i$, its posterior draws are obtained directly from the
fitted model, without a separate imputation step.
\newpage
\section{Stan code for the GP-GRM}
\label{suppl:stan_code}

Listing~\ref{lst:gp_grm} gives the Stan code of the GP-GRM. Variable names follow the notation of Section~\ref{sec:methodology}: \texttt{C}, \texttt{J}, \texttt{T}, and \texttt{I} denote the numbers of response categories, items, measurement occasions, and respondents; \texttt{alpha\_tilde}, \texttt{alpha}, and \texttt{tau} denote the raw log-discriminations $\tilde{\alpha}_j$, the discriminations $\alpha_j$, and the thresholds $\boldsymbol{\tau}_j$; \texttt{ell} denotes the length scales $\ell_i$; and \texttt{z}, \texttt{K}, \texttt{L}, and \texttt{theta} denote the standard-normal variables $\boldsymbol{z}_i$, the kernel matrix $\boldsymbol{K}_i$, its Cholesky factor $\boldsymbol{L}_i$, and the latent trajectories $\boldsymbol{\theta}_i$ of the non-centered parameterization (Section~\ref{sec:estimation_priors}). The data are supplied in long format, with one row per observed response.

\begin{lstlisting}[language=Stan, basicstyle=\footnotesize\ttfamily, columns=fullflexible, keepspaces=true, breaklines=true, caption={Stan code of the GP-GRM}, label={lst:gp_grm}]
data {
  // Number of response categories
  int<lower=2> C;
  // Total number of observed responses
  int<lower=1> N;
  // Number of items
  int<lower=1> J;
  // Number of measurement occasions
  int<lower=1> T;
  // Number of respondents
  int<lower=1> I;
  // Respondent index i of each response
  array[N] int<lower=1, upper=I> person;
  // Item index j of each response
  array[N] int<lower=1, upper=J> item;
  // Occasion index t of each response
  array[N] int<lower=1, upper=T> occasion;
  // Time points t_1, ..., t_T
  array[T] real time_points;
  // Observed responses
  array[N] int<lower=1, upper=C> y;
  // 1 = fill log_lik and y_rep (0 while sampling,
  // 1 in generate_quantities on a subset of draws)
  int<lower=0, upper=1> compute_gq;
}

transformed data {
  // Small jitter added to the main diagonal of the kernel for numerical stability
  real delta = 1e-5;

  // Hard bounds on the length scale: min / max time difference between waves
  real ell_min = 0.5;
  real ell_max = 12;
}

parameters {
  // --- Item parameters ---
  // Raw log-discriminations before centering
  vector[J] alpha_tilde;
  // Item-specific thresholds
  array[J] ordered[C - 1] tau;

  // --- GP parameters ---
  // Standard-normal variables (non-centered GP)
  matrix[I, T] z;
  // Length scale per respondent
  vector<lower=ell_min, upper=ell_max>[I] ell;
}

transformed parameters {
  // Centered discriminations: product of alpha = 1 (sum of log-discriminations = 0)
  vector<lower=0>[J] alpha = exp(alpha_tilde - mean(alpha_tilde));

  // Latent trajectories: rows = respondents (I), columns = occasions (T)
  matrix[I, T] theta;

  for (i in 1:I) {
    // Exponentiated quadratic kernel with unit marginal variance
    matrix[T, T] K = add_diag(gp_exp_quad_cov(time_points, 1, ell[i]), delta);

    // Cholesky decomposition for the non-centered parameterization: L * z ~ N(0, K)
    matrix[T, T] L = cholesky_decompose(K);
    theta[i] = (L * z[i]')';
  }
}

model {
  // --- Item priors ---
  for (j in 1:J) {
    // Thresholds: Fox (2010)
    tau[j] ~ normal(0, 4);
  }
  // Discriminations: McBride & Wang (2026); = lognormal(0, 0.5) on exp(alpha_tilde)
  alpha_tilde ~ normal(0, 0.5);

  // --- GP priors ---
  // Length scale: Betancourt (2020)
  // Truncated to [0.5, 12] by the bounds (constant normalization, no T[] needed)
  // -> p(ell<1) ~ 0.014, p(ell<2) ~ 0.25, median ~ 2.9
  ell ~ inv_gamma(3, 8);

  // Non-centered GP parameterization
  to_vector(z) ~ std_normal();

  // --- Likelihood ---
  for (n in 1:N) {
    real eta = alpha[item[n]] * theta[person[n], occasion[n]];
    y[n] ~ ordered_logistic(eta, tau[item[n]]);
  }
}

generated quantities {
  // Size zero unless compute_gq = 1 (keeps the sampling output small)
  // Pointwise log-likelihood
  vector[compute_gq ? N : 0] log_lik;
  // Posterior predictive draws
  array[compute_gq ? I : 0, J, T] int<lower=1, upper=C> y_rep;

  if (compute_gq) {
    // Pointwise log-likelihood
    for (n in 1:N) {
      real eta = alpha[item[n]] * theta[person[n], occasion[n]];
      log_lik[n] = ordered_logistic_lpmf(y[n] | eta, tau[item[n]]);
    }

    // Posterior predictive simulation
    for (i in 1:I) {
      for (j in 1:J) {
        for (t in 1:T) {
          real eta = alpha[j] * theta[i, t];
          y_rep[i, j, t] = ordered_logistic_rng(eta, tau[j]);
        }
      }
    }
  }
}
\end{lstlisting}

\clearpage

\printbibliography[heading=subbibliography,title={References for the Supplementary Material}]
\end{refsection}

\end{document}